\documentclass[showpacs,aps,prd,superscriptaddress,nofootinbib,floatfix,amsmath,amssymb]{revtex4}
\usepackage{graphicx}
\usepackage{amssymb}

\begin{document}


\title{ On-shell $WW\gamma$ vertex in the T-Parity and non T-Parity Littlest Higgs model}
\author{A. Moyotl}
\email[E-mail:]{amoyotl@sirio.ifuap.buap.mx}\affiliation{Instituto de F\'{\i}sica, Universidad
Aut\'onoma de Puebla, Apartado Postal J-48, 72570, Puebla, M\'exico}
\author{G. Tavares-Velasco}
\email[E-mail:]{gtv@fcfm.buap.mx} \affiliation{Facultad de
Ciencias F\'\i sico Matem\' aticas, Benem\' erita Universidad
Aut\' onoma de Puebla, Apartado Postal 1152, Puebla, Pue.,
M\'exico}

\date{\today}

\begin{abstract}

The  static electromagnetic properties of the $W$ boson, $\Delta
\kappa$ and $\Delta Q$, are calculated in the T-parity and non
T-parity littlest Higgs model (LHM) including terms up to the order of $(v/f)^4$,
with $v$ the standard model (SM) vacuum expectation value and $f$ the scale of the global symmetry breaking.  There are contributions  from the virtual effects of the new heavy particles and also from the new corrections to the SM vertices, which depend on the mixing parameter $c$ and decrease quickly as $f$ increases. Depending on the value of $c$, the partial contributions to $\Delta \kappa$ and $\Delta Q$ can add constructively or destructively. The terms of the order of
$(v/f)^4$ are subdominant but they can add constructively and can enhance the   $W$ form factors  for $f \lesssim 1$ TeV. In general the size of the $W$ form factors is very suppressed in the LHM without T-parity as the constraints on the scale $f$ from electroweak precision measurements are very tight. The LHM with T-parity has no such constraints and it allows for values of $W$ form factors similar to those found in other weakly coupled SM extensions, though they are out of the expected  sensitivity of future measurements at the LHC. We also discuss the possibility that there is some enhancement due to the interactions arising from the strongly interacting ultraviolet completion of the LHM.
\end{abstract}

\pacs{13.40.Gp, 12.60.Fr, 14.70.Fm}

\maketitle

\section{Introduction}
\label{introduction}

It has been long known that trilinear gauge boson couplings (TGBCs)
are very sensitive to new physics effects and  may serve as a probe
of the gauge sector of the standard model (SM) \cite{Hagiwara}. It
is expected that this class of couplings will be experimentally
tested with high accuracy at a near future \cite{Wudka}. Particular
interest has been put on the couplings involving the charged gauge
boson $W$, namely the $WW\gamma$ and $WWZ$ vertices, whose one-loop
corrections were long ago studied in the SM \cite {WWgSM,Argyres}
and afterwards in the framework of several of its extensions
\cite{Couture:1987eu,Bilchak:1986bt,Couture:1987xq,Lahanas:1994dv,Aliev:1985an,TavaresVelasco:2001vb,GarciaLuna:2003tj}.
The on-shell $WW\gamma$ vertex can be written in terms of four form factors that
define the CP-even and CP-odd static electromagnetic properties of
the $W$ boson \cite{Hagiwara}. There are two CP-odd form factors,
which are absent at the one-loop level in the SM and are thus
expected to be negligibly small. These  terms can only be generated
at the one-loop level in models in which the $W$ boson couples
simultaneously to both left- and right-handed fermions
\cite{Burgess}. As far as the CP-even static electromagnetic
properties are concerned, they are determined by two form factors,
$\Delta \kappa$ and $\Delta Q$ \cite{Hagiwara}, which can only arise
at the one-loop level in the SM and any other renormalizable theory,
thereby being highly sensitive to new physics effects. It is then interesting to assess the
deviation of these form factors from their SM values in any extended
model.

Among recently proposed SM extensions, little Higgs models
\cite{ArkaniHamed:2001nc,ArkaniHamed:2002qy,ArkaniHamed:2002pa,ArkaniHamed:2002qx,Gregoire:2002ra,Low:2002ws,Skiba:2003yf,Chang:2003un,Chang:2003zn,
Cheng:2004yc,Schmaltz:2004de,Kaplan:2003uc,Agashe:2004rs,Birkedal:2006fz,Low:2004xc}
have been the source of considerable interest. The parameters of
this class of models have been constrained from low energy
electroweak measurements
\cite{Csaki:2002qg,Hewett:2002px,Csaki:2003si,Gregoire:2003kr,Chen:2003fm,Casalbuoni:2003ft,Marandella:2005wd,Han:2005dz,
Kilian:2003xt, Hubisz:2005tx} and their phenomenological
consequences at future colliders have been examined
\cite{Burdman:2002ns, Han:2003wu, Azuelos:2004dm,  Han:2005ru,
Han:2003gf,
Ros:2004ag,Liu:2004pv,Yue:2006qx,Kilian:2006eh,Cheung:2008zu,Cho:2004xt,Kong:2007uu,Logan:2004hj,Huang:2009nv,Yue:2005av,Kai:2007ji,
Cheung:2006gf, Cheung:2006nk, Choudhury:2006sq, Yue:2007ym,
Hubisz:2004ft, Dib:2005re,  Chen:2006cs, Belyaev:2006jh,
Freitas:2006vy, Wang:2006ke, Yue:2006hn, Carena:2006jx,
Hundi:2006rh, Cao:2006wk, Matsumoto:2006ws, Choudhury:2006mp,
HongSheng:2007ve, Wang:2007zx, Cao:2007pv, Datta:2007xy,
Wang:2007nf, Cao:2008qd, Wang:2008qna, Wang:2008iw, Yue:2008zp,
Matsumoto:2008fq, Penunuri:2008pb, Wang:2008zw, Yue:2009cq}. It
turns out that this class of models predict interesting new physics
at the TeV scale that could  be tested at the large hadron collider
(LHC)
\cite{Han:2003wu,Azuelos:2004dm,Han:2005ru,Ros:2004ag,Liu:2004pv,Yue:2006qx,Kilian:2006eh,Cheung:2008zu,Matsumoto:2006ws,Choudhury:2006mp,Datta:2007xy,
Wang:2007nf, Wang:2008iw, Matsumoto:2008fq,Wang:2008zw} or the
planned international linear collider (ILC)
\cite{Cho:2004xt,Kong:2007uu,Logan:2004hj,Huang:2009nv,Yue:2005av}.
Little Higgs theories are appealing as they provide a solution to
the hierarchy problem different to the one offered by supersymmetry
or extra dimensions theories: by requiring the Higgs boson to be a
pseudo-Goldstone boson arising from a spontaneously broken
approximate symmetry at a $\Lambda_S$ scale, the quadratically
divergent SM one-loop contributions to the Higgs boson mass are
canceled without fine tuning up to the $\Lambda_S$ scale, thereby
rendering a naturally light Higgs boson \cite{ArkaniHamed:2001nc,
ArkaniHamed:2002qy, ArkaniHamed:2002pa}. In order to avoid fine
tuning, the Higgs boson couplings must also be fixed in a specific
way.  Little Higgs models predict new particles: fermions, gauge
bosons and scalar bosons,  with a mass of the order of the TeV
scale, which cancel the SM one-loop quadratic divergences to the
Higgs boson mass \cite{ArkaniHamed:2001nc, ArkaniHamed:2002qy,
ArkaniHamed:2002pa}. Even if these new particles are too heavy to be
directly produced at particle colliders, they may show-up through
loop effects in particle observables. In this work we are interested
in analyzing the one-loop contributions to $\Delta \kappa$ and
$\Delta Q$ in the framework of the most popular realization of
little Higgs models, which is the so called littlest Higgs model
(LHM) \cite{ArkaniHamed:2002qy}. In such a model there are new
contributions to the one-loop $WW\gamma$ vertex arising from
corrections to the SM couplings and also from the virtual effects of
the new particles. We will see below that these contributions can be
conveniently classified in a series of  powers of $v/f$, with $v$
the standard model vacuum expectation value and $f\sim O({\rm TeV})$
\cite{ArkaniHamed:2002qy} the scale of the global symmetry breaking.
We will present explicit analytical expressions for  $\Delta \kappa$
and $\Delta Q$,  and analyze their size for representative values of
the model parameters consistent with low energy constraints. We will
also examine the behavior of these form factors in the LHM with
T-parity, which is a more recent version of the LHM that has
attracted considerable interest recently: constraints on its
parameter space have been obtained \cite{Hubisz:2005tx} and its
phenomenology has been extensively studied \cite{Hubisz:2004ft,
Dib:2005re, Chen:2006cs, Belyaev:2006jh, Freitas:2006vy,
Wang:2006ke, Yue:2006hn, Carena:2006jx, Hundi:2006rh, Cao:2006wk,
Matsumoto:2006ws, Choudhury:2006mp, HongSheng:2007ve, Wang:2007zx,
Cao:2007pv, Datta:2007xy,Wang:2007nf, Cao:2008qd, Wang:2008qna,
Wang:2008iw, Yue:2008zp, Matsumoto:2008fq, Penunuri:2008pb,
Wang:2008zw, Yue:2009cq}. The addition of T-parity to the LHM is
meant to alleviate some of its problems. In particular, the
constraints on the scale $f$ from electroweak precision observables
are significantly relaxed \cite{Hubisz:2005tx}. Furthermore, an
exciting by-product of this model is a dark matter candidate
\cite{Low:2004xc}.

The rest of the paper is organized as follows. Sec. \ref{TheoFrame} is
devoted to a brief description of the LHM. The definition of the static electromagnetic properties of the $W$ boson is introduced in Sec. \ref{theoretical}, whereas in Sec. \ref{calculation} we
present the analytical expressions for the static properties of the $W$ boson. The
numerical results are analyzed in Sec. \ref{Analysis} and the
conclusions are presented in Sec. \ref{Remarks}.

\section{Overview of the littlest Higgs model}
\label{TheoFrame}

We now turn to present an overview of the LHM \cite{ArkaniHamed:2002qy} and also discuss  the inclusion of T-parity \cite{Low:2004xc}. We will content ourselves with discussing those topics relevant for our calculation. A detailed
description of these models along with their  Feynman rules  and an
analysis of their phenomenology can be
found in Refs. \cite{Han:2003wu} and \cite{Hubisz:2004ft,Chen:2006cs}. The LHM and the LHM with T-parity are almost identical in the scalar and gauge sectors, but their fermion sectors differ. We will discuss first those features shared by both models.


The LHM
\cite{ArkaniHamed:2002qy}, is a non-linear sigma model based on an $SU(5)$ global
symmetry, with a locally gauged $[SU(2) \otimes U(1)]_{1} \otimes
[SU(2) \otimes U(1)]_{2}$ symmetry as a subgroup. At the cutoff
scale $\Lambda_S$, the $SU(5)$ global symmetry is broken down to
$SO(5)$ by the vacuum expectation value (VEV) of the sigma field,
$\Sigma_0$, which is of the order of $f\sim O$(TeV). At this stage
the gauged symmetry is also broken down to its diagonal subgroup,
$SU(2) \times U(1)$, which happens to be the SM gauge group.
Fourteen Goldstone bosons arise after the breaking of the global
symmetry and four of them, which transform under the electroweak
gauge group as a real singlet and a real triplet, are eaten by the
gauge bosons associated with the broken gauge symmetry, which thus
acquire a mass of the order of $f$. The remaining ten Goldstone
bosons transform as a complex doublet $h$ and a complex triplet $\phi$ of the
electroweak gauge group. A Coleman-Weinberg  type potential for
these ten Goldstone bosons is induced radiatively by the gauge and
Yukawa couplings that break the global $SO(5)$ symmetry. The complex
triplet gets a mass of the order $f$, while the neutral component of
the complex doublet develops a VEV, $v$, that is responsible for the
electroweak symmetry breaking (EWSB). In addition, it is necessary
to introduce a new vector-like top quark to cancel the quadratically
divergent contribution to the Higgs mass from the top quark loop.

The Lagrangian of the LHM can be written as the sum of the kinetic
energy Lagrangian of the $\Sigma$ field, $\mathcal{L}_{\rm K}$, plus
the Yukawa Lagrangian, $\mathcal{L}_{\rm Y}$, and also the kinetic terms of the gauge and fermion sectors.  The sigma field kinetic Lagrangian is given by
\cite{ArkaniHamed:2002qy}
\begin{equation}
\mathcal{L}_{\rm K}= \frac{f^{2}}{8} {\rm Tr} | D_{\mu} \Sigma |^2,
\label{kinlag}
\end{equation}
with the $[SU(2)\times U(1)]^2$ covariant derivative defined by \cite{ArkaniHamed:2002qy}
\begin{equation}
D_{\mu} \Sigma = \partial_{\mu} \Sigma
- i \sum_{j=1}^2 \left[ g_{j} W_{j\,\mu}^{a} (Q_{j}^{a}\Sigma + \Sigma Q_{j}^{a\,T})
+ g'_{j} B_{j\,\mu} (Y_{j} \Sigma+\Sigma Y_{j}^{T}) \right].
\end{equation}
The heavy $SU(2)$ and $U(1)$ gauge bosons are  $W_{j}^\mu =
\sum_{a=1}^{3} W_{j}^{\mu \, a} Q_{j}^{a}$ and $B_{j}^\mu =
B_{j}^{\mu} Y_{j}$, with  $Q_j^a$ and $Y_j$ the gauge generators,
while $g_i$ and $g'_i$ are the respective gauge couplings. The VEV
$\Sigma_0$ generates masses for the gauge bosons of
the order of $f$ and mixings between them. The heavy gauge boson mass eigenstates are given
by \cite{ArkaniHamed:2002qy}
\begin{eqnarray}
W'^a &=& -c W_{1}^a + s W_{2}^a,\\
B' &=& -c^{\prime} B_{1} + s' B_{2}  ,
\end{eqnarray}
with masses  $m_{W'} = \frac{f}{2}\sqrt{g_1^2+g_2^2}$ and
$m_{B'} = \frac{f}{\sqrt{20}} \sqrt{g'^2_{1}+g'^2_{2}}$.

The orthogonal combinations of gauge bosons are identified as the SM
gauge bosons \cite{ArkaniHamed:2002qy}:
\begin{eqnarray}
W^a &=& s W_{1}^a + c W_{2}^a,\\
B &=& s' B_{1} + c' B_{2},
\end{eqnarray}
which remain massless at this stage,
their couplings being given by
$g = g_{1}s = g_{2}c$ and
$g' = g'_{1}s' = g'_{2}c'$,
where $s = g_{2}/\sqrt{g_{1}^{2}+g_{2}^{2}}$ and
$s' = g'_{2}/\sqrt{g'^2_{1}+g'^2_{2}}$ are mixing parameters (here $c=\sqrt{1-s^{2}}$ and $c' = \sqrt{1-s'^{2}}$).

After EWSB there is additional mixing between heavy and light gauge
bosons and we arrive at the final mass eigenstates. There are three
light gauge bosons (the SM ones) $A_L$, $W_L$ and $Z_L$, and three
heavy gauge bosons which are their counterpart;  $A_H$, $W_H$ and
$Z_H$. One of the light gauge bosons is the photon, $A_L$, which
remains massless, but the $W_L$ and $Z_L$ gauge bosons masses get
corrected by terms of the order of $(v/f)^2$ and so are the masses
of the heavy gauge bosons. The latter obey the following relations
\cite{Han:2003wu}:

\begin{equation}
m_{Z_H}^2\simeq m_{W_H}^2=m_W^2\left(\frac{f^2}{s^2c^2v^2}-1\right)\ge 4 m_W^2 \frac{f^2}{v^2}, \label{WHmasslimit}\\
\end{equation}
\begin{equation}
m_{A_H}^2= m_Z^2 s_W^2\left(\frac{f^2}{5 s'^2c'^2v^2}-1+\frac{x_H c_W^2}{4s^2 c^2 s_W^2}\right)\ge 4 m_W^2 t_W^2 \frac{f^2}{5v^2}, \label{AHmasslimit}
\end{equation}
with $t_W=s_W/c_W$, being $s_W$ and $c_W$ the cosine and sine of the Weinberg angle $\theta_W$, while $x_H=\frac{5}{2}gg'\frac{scs'c'(c^2s'^2+s^2c'^2)}{5 g^2s'^2c'^2-g'^2s^2c^2}$.

In the scalar sector, a Coleman-Weinberg potential $V_{\rm CW}$, which  is generated radiatively at one loop by
the heavy gauge bosons and the top-partner, induces the EWSB. After expanding the $\Sigma$ field, the following  $V_{CW}$
potential is obtained \cite{ArkaniHamed:2002qy}
\begin{equation}
    V_{\rm CW} = \lambda_{\phi^2} f^2 {\rm Tr}|\phi|^2
    + i \lambda_{h \phi h} f \left( h \phi^\dagger h^T
        - h^* \phi h^\dagger \right)
    - \mu^2 |h|^2
    + \lambda_{h^4}  |h|^4,
\end{equation}
where $\lambda_{\phi^2}$, $\lambda_{h \phi h}$, and $\lambda_{h^4}$
depend on the fundamental parameters of the model. As for $\mu^2$, which receives logarithmic divergent
contributions at one-loop level and quadratically divergent
contributions at the two-loop level, is treated as a free parameter
of the order of $f^2/16 \pi^2$.

Minimizing the $V_{\rm CW}$ potential leads to the VEVs
of the SM Higgs doublet and the $SU(2)$ Higgs triplet, $v$ and $v'$.
After diagonalizing the Higgs mass matrix, the light Higgs boson
mass can be obtained at the leading order \cite{Han:2003wu}
\begin{equation}
    m^2_{H}= 2 \mu^2 = 2 \left( \lambda_{h^4}
    - \frac{\lambda_{h \phi h}^2}{ \lambda_{\phi^2}} \right) v^2 .
\end{equation}

It is required that  $\lambda_{h^4} > \lambda_{h \phi h}^2 /
\lambda_{\phi^2}$ to obtain the correct  electroweak symmetry
breaking vacuum with $m^2_H>0$. The Higgs triplet masses are
degenerate at this order and can be written as follows
\cite{Han:2003wu}:
\begin{equation}
 m_\Phi^2=\frac{2 m_H^2 f^2}{v^2\left(1-\frac{16v'^2f^2}{v^4}\right)}.
\end{equation}
To have a positive definite  $m_\Phi^2$ it is necessary that

\begin{equation}
\frac{v'^2}{v^2} < \frac{v^2}{16f^2}. \label{boundv'/v}
\end{equation}
The mass of the Higgs triplet obeys thus the following relation
\begin{equation}
 m_\Phi^2\ge 2m_H^2\frac{f^2}{v^2}. \label{Phimasslimit}
\end{equation}

In summary, in the gauge sector there are four new gauge bosons $W_H^\pm$, $Z_H$
and $A_H$, while in the scalar sector there are new neutral, singly charged and doubly charged Higgs
scalars, $\phi^0$, $\phi^-$, $\phi^{--}$, together with one
pseudoscalar boson $\phi^P$. The presence of the heavy gauge bosons
and the heavy top quark partner is generic in little
Higgs models since these particles are necessary for the collective symmetry
breaking \cite{Perelstein:2005ka}. However, the scalar sector depends on the particular implementation of the model.


In the fermion sector  it is necessary to introduce a new
vector-like top quark $T$, dubbed the top partner. The $T$ loops
cancel the quadratically divergent contribution to the Higgs mass arising
from the top quark loops. This fixes the Yukawa interactions, given
by \cite{ArkaniHamed:2002qy}

\begin{equation}
\mathcal{L}_{\rm Y} = \frac{1}{2}
\lambda_{1} f \epsilon_{ijk} \epsilon_{x,y} \chi_{i}
\Sigma_{j,x}\Sigma_{k,y} {u'}_{3}^{c}
+ \lambda_{2} f \tilde{t} \tilde{t}^{c} + {\rm H.c.},
\end{equation}
where $t_3$ is the SM top quark, $u'_{3}$ is the SM right-handed top
quark, $({\tilde t},\tilde{t'}^{c})$ is a new vector-like top quark
and ${\chi}=(b_{3}, t_{3},  \tilde{t})$. The first term of
${\mathcal L}_{\rm Y}$ induces the couplings of the Higgs boson to
the fermions such that the quadratic
divergences from the top quark loop are canceled by the top partner loop.
The expansion of the $\Sigma$ field leads to the physical states,
$t$ and $T$, after diagonalizing the mass matrix. At the leading
order in $v/f$, the masses of the SM top quark and the new top quark $T$ are given by
\cite{Han:2003wu}

\begin{equation}
    m_t = \frac{\lambda_1 \lambda_2}{\sqrt{\lambda_1^2 + \lambda_2^2}} v,\quad
    m_T = f \sqrt{\lambda_1^2 + \lambda_2^2}.\label{Topmass}
\end{equation}

There is no need to introduce extra vector-like quarks for the first two quark generations as the quadratic divergences from the first two families of fermion loops are not important below the cutoff scale $\Lambda_S$.

\subsection{LHM with T-parity}

The LHM can give some large corrections to electroweak precision observables, so stringent constraints on the scale $f$ are imposed by experimental data. An interesting solution to this problem is obtained by invoking a discrete $Z_2$ symmetry, T-parity, which can eliminate such dangerous corrections to electroweak precision observables and relax the constraints on $f$ \cite{Low:2004xc}.
T-parity is a symmetry analogue to the R-parity symmetry introduced in supersymmetric models. In this
realization of the LHM, known as the LHM with T-parity, the SM particles are T-even whereas the
heavy particles are T-odd, thereby forbidding any effective operator
involving only light fields to be generated at the tree-level by
exchange of heavy fields since an even number of these fields is
required at each vertex. This prevents dangerous corrections to
electroweak precision observables \cite{Low:2004xc}.

The gauge and scalar sectors of the LHM with T-parity  are identical
to the original LHM with the restrictions $g_1 = g_2$ and $g'_1 =
g'_2$ \cite{Low:2004xc}. It means that in the gauge sector the
T-parity transformation just exchanges the $[SU(2) \times U(1)]_1$
and $[SU(2) \times U(1)]_2$ gauge bosons. The light SM gauge bosons
are T-even while the heavy gauge bosons are T-odd. In the scalar
sector, the SM Higgs doublet is T-even, while the additional
$SU(2)_L$ triplet $\Phi$ is T-odd. The $H\Phi H$ coupling is thus forbidden and so is
a nonzero VEV, $v'$, for the $SU(2)_L$ triplet.  As a result, in the LHM with T-parity the following conditions are automatically satisfied:

\begin{eqnarray}
\label{RepTPar1}
c&=&s,\quad c'=s', \\
s_0&=&s_P=s_+=0,\\
\label{RepTPar2} c_0&=&c_P=c_+=1. \label{RepTPar3}
\end{eqnarray}

In
summary, in the gauge and scalar sectors both the LHM and the LHM
with T-parity have the same particle content \cite{Low:2004xc}. The relations for the
masses of the heavy gauge bosons given in Eqs. (\ref{WHmasslimit}) and (\ref{AHmasslimit})
are valid in the LHM with T-parity and so is the relation
(\ref{Phimasslimit}) for the mass of the Higgs triplet.

As far as the fermion sector is concerned,
T-parity is implemented in such a way that all the SM fermions are T-even. While in the LHM only the
third quark generation is modified, the introduction of T-parity
requires the doublet spectrum to be doubled to avoid compositeness
constraints \cite{Low:2004xc}. For each SM fermion doublet, two fermion doublets are
introduced. Further details of the fermion sector of the LHM with T-parity can be found elsewhere \cite{Hubisz:2004ft}. For the purpose of our calculation we only consider the top sector, which is additionally modified to cancel the quadratic divergences from the SM top quark loops. In addition to the SM top quark, whose mass is given by Eq. (\ref{Topmass}), there are two new mass eigenstates, namely, a T-even top partner $t'_+$ and a T-odd $t'_-$ top partner, with masses of the order of $f$ \cite{Hubisz:2004ft}:
\begin{equation}
    m_{t'_+} = f \sqrt{\lambda_1^2 + \lambda_2^2},\quad
    m_{t'_-} = \lambda_2f.
\end{equation}
The T-even top partner $t'+$ is counterpart of the top partner of the LHM, although its phenomenology is rather different. The new Yukawa interactions required to cancel the quadratic divergences of the top quark will correct some SM couplings with terms of the order of $(v/f)^2$. The $WW\gamma$ vertex will also be modified via the new ${W^+}\bar t'_+ b$ interaction, but there is no new contribution arising directly from the T-odd partner since by the T-parity symmetry it does not couple to the  $W^+b^-$ pair.

We now turn to present our calculation. The relevant Feynman rules in both the LHM and the LHM with T-parity can be found in
Ref. \cite{Han:2003wu}.  For completeness they are included in
Table \ref{Tab_Feynrules} and Table \ref{Tab_FeynrulesTP} of appendix \ref{FeynRules}.

\section{Static electromagnetic properties of the $W$ boson}
\label{theoretical}

The most general CP-conserving $WW\gamma$ vertex function with
on-shell particles can be written in terms of two form factors,
$\Delta \kappa$ and $\Delta Q$ \cite{Hagiwara}.  It is convenient to use the convention of Ref. \cite{WWgSM}
for the momenta of the external particles: $(p-Q)_\alpha$ and
$(p+Q)_\beta$ are the momenta of the incoming and outgoing $W^+$
gauge bosons and $-2Q_\mu$ is that of the emitted photon. In this
notation, the CP-even on-shell $WW\gamma$ vertex can be written as
\cite{WWgSM}:

\begin{equation}
\Gamma^{\mu \alpha \beta}=i\,e\left\{A\left[2\,p^\mu g^{\alpha
\beta}+4\,\left(Q^\beta\,g^{\mu \alpha}-Q^\alpha\,g^{\mu
\beta}\right)\right] +\,2\Delta \kappa \left(Q^\beta\,g^{\mu
\alpha}-Q^\alpha\,g^{\mu \beta}\right)+\frac{4 \,\Delta Q
}{m_W^2}p^\mu \,Q^\alpha\,Q^\beta\right\}.\label{WWg}
\end{equation}
In the SM $A=1$ at the tree-level, while both $\Delta \kappa$ and $\Delta
Q$ can only arise up to the one-loop level in renormalizable
theories, thereby being free from ultraviolet
singularities. These form factors can receive contributions from
fermions, gauge bosons, and scalar bosons that couple to the $W$
boson. The magnetic dipole moment ($\mu_W$) and the electric
quadrupole moment ($Q_W$) of the $W$ boson are defined in terms of
$\Delta \kappa$ and $\Delta Q$ as follows:

\begin{eqnarray}
\mu_W&=&\frac{e}{2\,m_W}\,\left(2+\Delta \kappa \right),\\
Q_W&=&-\frac{e}{m_W^2}\,\left(1+\Delta \kappa +\Delta Q\right).
\end{eqnarray}

It is worth mentioning that another parametrization is used for the $WW\gamma$ vertex in experimental works. We choose to use the parametrization of Eq. (\ref{WWg})  as it is the one that has been used in the theoretical calculations of  the one-loop $WW\gamma$ vertex.

\subsection{Experimental limits on the electromagnetic $W$ form factors}

We now would like to comment on the current limits on the $W$ form factors and the expected sensitivity of future measurements at the LHC and the planned ILC. In experimental works studying the  limits on trilinear gauge boson couplings, it is usual to introduce a different parametrization for the $WW\gamma$ vertex, for which the following Lagrangian is considered \cite{Hagiwara}:

\begin{equation}
{\cal L}_{WW\gamma}=-ie\left(g_1^\gamma\left(W^+_{\mu \nu}W^{\mu}A^{\nu}-W^+_{\mu}A_\nu W^{\mu\nu}\right)+\kappa_\gamma W^+_{\mu}W_\nu A^{\mu\nu}+\frac{\lambda_\gamma}{m_W^2}W^+_{\lambda \mu} W^\mu_\nu A^{\nu\lambda}\right)
\label{WWgHagi}
\end{equation}
with $V^{\mu\nu}=\partial^\mu V^\nu-\partial^\nu V^\mu$. In this parametrization  $\Delta\kappa=\kappa_\gamma-1+\lambda_\gamma$ and $\Delta Q=-2\lambda_\gamma$. In the SM $g_1^\gamma=\kappa_\gamma=1$ and $\lambda_\gamma=0$ at the tree level. Furthermore, it is usual to introduce the definition $\Delta \kappa_\gamma\equiv \kappa_\gamma-1$, which evidently  is not the same as the $\Delta\kappa$ form factor used throughout this work.

The most stringent limits on trilinear gauge boson couplings up to date were obtained by the Delphi collaboration
using the data from $W^+W^-$ and $W$ production at the large electron positron collider (LEP2) at centre-of-mass energies between 189 and 209 GeV \cite{:2010zj}. The data from the $jj\ell\nu$, $jjjj$, $jjX$ and $\ell X$ final states were used, where $j$, $\ell$ and $X$ represent a jet, a lepton and missing four-momentum, respectively. A fit to a single parameter while keeping held the other parameters places the following limits \cite{:2010zj}:
\begin{equation}
\lambda_\gamma= 0.002^{+0.035}_{-0.035}
\end{equation}
\begin{equation}
\Delta \kappa_\gamma= 0.024^{+0.077}_{-0.081}
\end{equation}

The limits obtained at Tevatron by the D0 collaboration by combining measurements of diboson production in $p\bar p$ collisions at  $\sqrt s = 1.96$ TeV are less stringent \cite{Abazov:2009tr} but it is expected that they reach the level of the LEP2 limits when combined with the CDF data \cite{Abazov:2009tr}. Although the LEP and Tevatron data are consistent with the SM, new physics effects cannot be ruled out. However, to disentangle any new physics from the SM radiative corrections it would be necessary to  achieve a very high experimental sensitivity to anomalous trilinear gauge boson couplings. While the sensitivity to $\Delta\kappa_\gamma$ and $\lambda_\gamma$ is above the $10^{-2}$ level at the Tevatron, the expected sensitivity at the LHC via $pp \to W\gamma$ is  $|\Delta\kappa_\gamma|\sim 5\times 10^{-3}$ and $|\lambda_\gamma|\sim 3\times 10^{-4}$, with an integrated luminosity of 300 fb$^{-1}$ and a running of three years \cite{Djouadi:2007ik}. The sensitivity to $\Delta\kappa_\gamma$ might be considerably improved at the future ILC via $e^-e^+\to W^-W^+$.  The projected sensitivity of this machine running at 800  GeV with an integrated luminosity of 1000 fb$^{-1}$ is $|\Delta\kappa_\gamma|\sim 10^{-4}$ and $|\lambda_\gamma|\sim 2\times 10^{-4}$ \cite{Djouadi:2007ik}. Three years of running are also assumed.

\section{The $W$ form factors in the LHM with and without T-parity}
\label{calculation}
We now turn to present the analytical expressions for $\Delta\kappa$
and $\Delta Q$ in the framework of the LHM, and the corresponding
results for the LHM with T-parity will follow straightforwardly.  Our
calculation is done in the unitary gauge and Feynman parametrization
is used for the loop integrals. Before presenting our calculation we would like to make some remarks about the order of the distinct contributions to the $W$ form factors. We will consider the following types of corrections:

\begin{itemize}
\item Type a: these are contributions arising from loops including heavy particles (with mass $M\sim f$). A closer study of the loop amplitudes shows that this class of contributions  behave as $(m_W/M)^2\sim (v/f)^2$ for large $f$. This is also observed in contributions to the $W$ form factors from other weakly coupled theories \cite{TavaresVelasco:2001vb}.

\item Type b: these are  contributions that arise from loops including only SM particles when the LHM corrections to the SM vertices are considered.  It turns out that these contributions  are proportional to $(v/f)^2$ and are expected to be of similar size to the ones of type a.

\item Type c: some LHM couplings are zero at the lowest order and arise up to higher orders of $v/f$. Extra contributions to the $W$ form factors can  arise from loops with heavy particles when the higher order corrections to the LHM couplings are considered.  These contributions have a suppression factor of $(v/f)^2$ due to the loop amplitude and an additional suppression coming from the involved couplings. These contributions are of the order of $(v/f)^4$ and they can be important only for $f \lesssim 1$ TeV but they will be negligible for larger $f$
\end{itemize}

\subsection{Fermion  contributions}

\begin{figure}
 \centering
\includegraphics[width=2in]{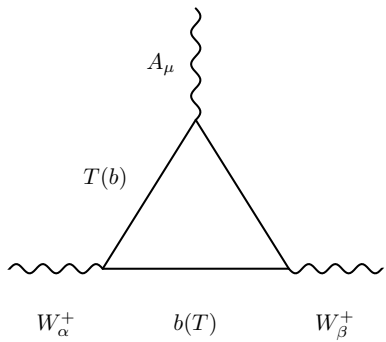}
\caption{Feynman diagrams for the fermion contribution to the
$WW\gamma$ vertex at the one-loop level in the LHM.   \label{fig_fermdiag}}
\end{figure}

There are  type a and type c contributions from $t$ and $T$ loops.
Since the SM coupling $Wbt$
is corrected by terms of the order of $(v/f)^2$, the $WW\gamma$ vertex including only SM quarks in the loop will receive contributions of the order of $(v/f)^2$. There are also contributions from loops with a top
partner $T$ and a $b$ quark, as shown in Fig. \ref{fig_fermdiag}. Since the $WTb$ coupling arises  at the order of $(v/f)$, the respective contribution to the $W$ form factors has an additional suppression factor of $(v/f)^2$ as this vertex enters twice in the loop amplitude. For the reasons explained before, these corrections can be important in the LHM with T-parity and we  will include them in our calculation.

The fermion triangle has been calculated previously \cite{Argyres} but we have recalculated this contribution. We will present the most general expressions for the $W$ form factors coming from type a and type c contributions, which are valid in both the LHM and the LHM with T-parity.

The fermion contributions to $\Delta\kappa$ can be written as
\begin{equation}
\Delta \kappa^{F}=\frac{g^2}{16 \pi^2}\frac{3v^2|V_{tb}|^2}{2f^2}\sum_{u=t,T}\int^1_0 a_u\left( Q_bf^{u,b}(x)-Q_u f^{b,u}(x)\right)dx,
\label{DeltaKF}
\end{equation}
with $a_u=-x_L^2+c^2(c^2-s^2)$ for $u=t$ and $a_u=x_L^2$ for $u=T$. Also
\begin{equation}
f^{i,j}(x)=\frac{x-1 }{\lambda^{i,j}}\left( -5\,x^2 + 3\,x^3 - x_j - \lambda^{i,j} + x\,\left( 2 + x_j + 3\,\lambda^{i,j} \right)  +
        \left(3\,x-1 \right) \,\lambda^{i,j}\,\log (\lambda^{i,j}) \right),
\end{equation}
where we introduced the shorthand notation $x_i=(m_j/m_W)^2$, while $\lambda^{i,j}$ depends on $x$:
\begin{equation}
\lambda^{i,j}(x)=x^2+(x_i-x_j-1)x+x_j.
\label{lambda}
\end{equation}

As far as the contribution to $\Delta Q$ is concerned, it is given by
\begin{equation}
\Delta Q^{F}=\frac{g^2}{16 \pi^2}\frac{3v^2|V_{tb}|^2}{2f^2}\sum_{u=t,T}\int^1_0 a_u\left( Q_bg^{u,b}(x)-Q_u g^{b,u}(x)\right)dx,
\label{DeltaQF}
\end{equation}
with
\begin{equation}
g^{i,j}(x)=\frac{4}{3\,\lambda^{i,j} }\left( 1 - x \right)^3x.
\end{equation}
In the above equations $x_L=\lambda_1^2/(\lambda_1^2+\lambda_2^2)$.
To get the contributions in the LHM with T-parity we must set $c=s$ and make the following replacement:  $x_L\to c_\lambda^2$ and $T\to t'_{+}$, with $c_\lambda=1-m_{t'_-}^2/m_{t'_+}^2$.

\subsection{Scalar  contributions of type a}

\begin{figure}
 \centering
\includegraphics[width=1.3in]{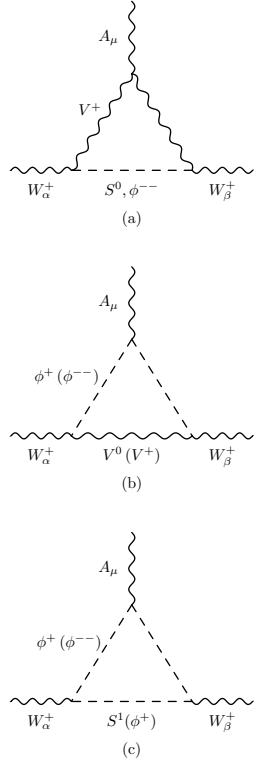}
\caption{Feynman diagrams for the scalar boson contribution  to the
on-shell $WW\gamma$ vertex. The
following notation is employed: $V^+$ stands for the charged gauge
bosons $W$ or $W_H$,  $S^0$ holds for the physical neutral scalar
bosons $H$ or $\phi^0$, $S^1$ can be $S^0$, $\phi^P$ or $\phi^{--}$,
whereas $V^0$ represents the massive neutral gauge bosons $Z$,
$Z_H$ or $A_H$.  In the LHM with T-parity, the scalar contributions arise from the Feynman diagram (a) with the
$W$ gauge boson and the Higgs boson $H$ in the loop,  and from the Feynman diagram (c).\label{fig_scaldiag}}
\end{figure}

There are  type a contributions from all the physical
scalar bosons  through 15 triangle
diagrams involving scalar particles and gauge bosons. All these contributions can be obtained from only three generic
Feynman diagrams, as depicted in Fig. \ref{fig_scaldiag}. Each one of these triangle diagrams
is free of ultraviolet singularities and renders an electromagnetic
gauge invariant amplitude by itself.

The type a corrections induced by the
Feynman diagrams of Fig. \ref{fig_scaldiag} can be written as

\begin{equation}
\Delta \kappa^{S}=\frac{g^2}{16 \pi^2}\sum_{i=a,b,c}\int^1_0 F_i(x) dx,
\label{DeltaKS}
\end{equation}
with
\begin{equation}
F_a(x)= \sum_{S=H,\phi^0}  h^{S,W_H} f_a^{S,{W_H}}(x) +h^{\phi^0,W} f_a^{\phi^0,{W}}(x)-\sum_{V=W,W_H}h^{\phi^{--},V} f_a^{\phi^{--},V}(x) ,
\label{F_a}
\end{equation}

\begin{equation}
F_b(x)= \sum_{V=Z,Z_H,A_H}  h^{V,\phi^-} f_b^{V,\phi^{-}}(x)+\sum_{V=W,W_H}  h^{V,\phi^{--}} f_b^{V,\phi^{--}}(x),
\end{equation}
and
\begin{equation}
F_c(x)=\sum_{S=H,{\phi^0},\phi^P} h^{S,\phi^-}
f_c^{S,\phi^{-}}(x)-h^{\phi^{--},\phi^-}
f_c^{\phi^{--},\phi^-}(x)+h^{\phi^-,\phi^{--}}
f_c^{\phi^{-},\phi^{--}}(x),\label{F_c}
\end{equation}
where the $h^{X,Y}$ coefficients, which account for all the coupling constants involved in each loop amplitude, can be written as
\begin{equation}
h^{X,Y}=\frac{\left|g_{XXA}\,g^2_{WXY}\right|}{e g^2}. \label{hXiXj}
\end{equation}
The couplings $g_{XXA}$ and $g_{WXY}$ can be obtained from Table
\ref{Tab_Feynrules} as they are the factors that multiply the
Lorentz parts $R_\mu$, $g_{\mu\nu}$, $S_{\mu\nu\rho}$ and
$T_{\mu\nu\rho\sigma}$ in the respective interaction vertices $XXA$
and $WXY$. The notation employed is evident since the external $W$
bosons couple to the internal particles $X$ and $Y$, whereas the
photon is emitted off the $Y$ particle. The explicit form of each
$h^{X,Y}$ coefficient is given in Table \ref{Tab_hXiXj}, while the
$f_i^{X,Y}(x)$ functions are given by

\begin{eqnarray}
f_a^{S,V}(x)&=&\frac{(1-x)^2(x^2+\lambda^{S,V}+3x_V)}{2m_W^2\lambda^{S,V}x_V},
\label{f_a}                                            \\
f_b^{V,S}(x)&=&-\frac{(1-x)\big(4x_Vx+\lambda^{V,S}(1-3x)\log(\lambda^{V,S})\big)}{2m_W^2\lambda^{V,S}x_V},
\label{f_b}                                           \\
f_c^{S_1,S_2}(x)&=&-2(3x^2-4x+1)\log(\lambda^{S_1,S_2}), \label{f_c}
\end{eqnarray}
and $\lambda^{X,Y}$ is given by Eq. (\ref{lambda}). Note that
the $f_i^{X,Y}(x)$ functions arise from the Feynman diagram labeled
(a), (b), and (c) in Fig. \ref{fig_scaldiag}. Therefore, the
partial contributions to $\Delta \kappa$ can be straightforwardly
obtained from the above equations.

\begin{table}[h]
\begin{tabular*}{0.3\textwidth}{@{\extracolsep{\fill}}lll}
\hline
$X$&$Y$&$\sqrt{h^{X,Y}}$\\
\hline
${H}$&${W_H}$&$\frac{g(c^2-s^2)v}{4sc}$\\
${\phi^0}$&${W_H}$&$\frac{g(c^2-s^2)(s_0 v-2\sqrt{2}v')}{4sc }$\\
${\phi^{--}}$&${W_H}$&$\frac{g(c^2-s^2)v'}{sc}$\\
${\phi^{0}}$&${W}$&$\frac{g(s_0v-2\sqrt{2}v')}{2}$\\
${\phi^{--}}$&${W}$&$2gv'$\\
${Z}$&${\phi^{-}}$&$g'v'$\\
${Z_H}$&${\phi^{-}}$&$\frac{g(c^2-s^2)v'}{2sc}$\\
${A_H}$&${\phi^{-}}$&$\frac{g'(c'^2-s'^2)(v{s_+}-4v')}{4gs'c'}$\\
${H}$&${\phi^{-}}$&$\frac{(\sqrt{2}s_0-s_+)}{2}$\\
${\phi^0}$&${\phi^{-}}$&$\frac{1}{\sqrt{2}}$\\
${\phi^P}$&${\phi^{-}}$&$\frac{1}{\sqrt{2}}$\\
${\phi^{--}}$&${\phi^{-}}$&$1$\\
\hline
\end{tabular*}
\caption{Constant coefficients $h_i^{X,Y}$ as given by Eq.
(\ref{hXiXj}). For the coefficients not listed,
$h^{X,\phi^{--}}=2h^{\phi^{--},X}$ for $X=W, W_H$, and $\phi^-$.
\label{Tab_hXiXj}}
\end{table}

As for the type a scalar contribution to $\Delta Q$, it is given by:

\begin{equation}
\Delta Q^{S}=\frac{g^2}{16 \pi^2}\sum_{i=a,b,c}\int^1_0 G_i(x) dx,
\label{DeltaQS}
\end{equation}
with
\begin{equation}
G_a(x)= \sum_{S=H,\phi^0}  h^{S,W_H} g_a^{S,{W_H}}(x) +h^{\phi^0,W} g_a^{\phi^0,{W}}(x)-\sum_{V=W,W_H}h^{\phi^{--},V} g_a^{\phi^{--},V}(x) ,
\label{G_a}
\end{equation}

\begin{equation}
G_b(x)= \sum_{V=Z,Z_H,A_H}  h^{V,\phi^-} g_b^{V,\phi^{-}}(x)+\sum_{V=W,W_H}  h^{V,\phi^{--}} g_b^{V,\phi^{--}}(x),
\end{equation}
and
\begin{equation}
G_c(x)=\sum_{S=H,{\phi^0},\phi^P} h^{S,\phi^-}
g_c^{S,\phi^{-}}(x)-h^{\phi^{--},\phi^-}
g_c^{\phi^{--},\phi^-}(x)+h^{\phi^-,\phi^{--}}
g_c^{\phi^{-},\phi^{--}}(x),\label{G_c}
\end{equation}
with the
$g_i^{X,Y}$ functions  given by

\begin{eqnarray}
g_a^{S,V}(x)&=&\frac{(1-x)^3x}{3m_W^2\lambda^{S,V}x_V},
\label{g_a}                                            \\
g_b^{S,V}(x)&=&\frac{(1-x)^3x}{3m_W^2\lambda^{V,S}x_V},
   \label{g_b}                                         \\
g_c^{S_1,S_2}(x)&=&\frac{4(1-x)^3x}{3\lambda^{S_1,S_2}}. \label{g_c}
\end{eqnarray}

Note that we have considered the most general scenario with nondegenerate masses for the heavy scalar triplet and the heavy gauge bosons $W_H$ and $Z_H$. In the degenerate mass approximation for the heavy scalar bosons, the contribution of the Feynman diagrams of type (c) where the photon is emitted off the singly charged scalar $\phi^-$ cancel, surviving only the Feynman diagram where the photon is emitted off the doubly charged scalar $\phi^{--}$.

\subsection{Scalar contributions of type b}

 These contributions arise from loops involving only SM particles. The only contribution is that given by the Feynman diagram \ref{fig_scaldiag}(a) with the SM gauge boson $W$ and the Higgs boson $H$ circulating in the loop. The respective contributions are
\begin{equation}
\Delta \kappa^{S}=\frac{g^2}{16 \pi^2}h^{H,W}\left(\frac{v}{f}\right)^2\int^1_0 f_a^{H,W}(x) dx,
\label{DeltaKSSM}
\end{equation}
and
\begin{equation}
\Delta Q^{S}=\frac{g^2}{16 \pi^2}h^{H,W}\left(\frac{v}{f}\right)^2\int^1_0 g_a^{H,W}(x) dx,
\label{DeltaQSSM}
\end{equation}
with $h^{H,W}=-\frac{2}{3}+(c^2-s^2)^2$.

 We have given the contribution to the $W$ form factors in the LHM. The respective contributions in LHM with T-parity are obtained when the relations given in Eqs. (\ref{RepTPar1})-(\ref{RepTPar3}) are considered. It means that apart from the contributions (\ref{DeltaKSSM}) and (\ref{DeltaQSSM}),  in the LHM with T-parity the $W$ form factors receive contributions from the Feynman diagrams of Fig. \ref{fig_scaldiag}(c) with only heavy sacalars.

\subsection{Scalar contributions of type c}

\begin{figure}
 \centering
\includegraphics[width=3in]{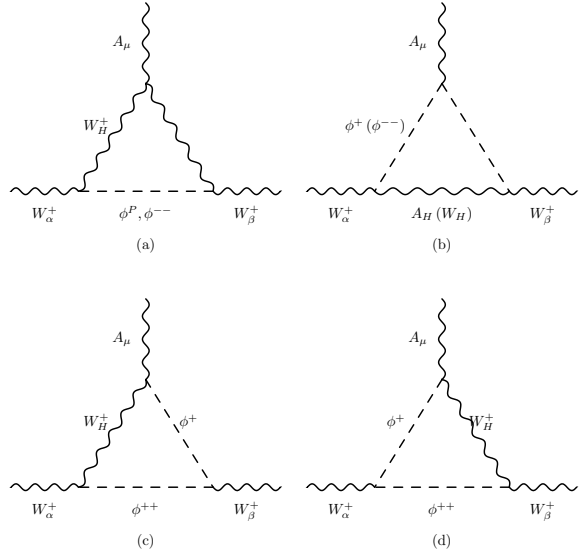}
\caption{Feynman diagrams that give contributions of type c to the $W$ form factors in the scalar sector of the LHM. \label{fig_scaldiagTP}}
\end{figure}

 This class of contributions comes from the four Feynman diagrams shown in Fig. \ref{fig_scaldiagTP}. Although this class of contributions are also present in the LHM without T-parity, they can be subdominant as in such a model the scale $f$ is expected to be much larger than $v$ due to electroweak constraints \cite{Csaki:2002qg}. We expect this contributions to be important only for the LHM with T-parity since in this model a value of $f=500$ GeV  is still consistent with electroweak precision mesurements \cite{Hubisz:2005tx}.

The higher-order contributions  to the $\Delta \kappa$ form factor  can be written as

\begin{equation}
\Delta \kappa^{S-H0}=\frac{g^2}{16\pi^2}\left(\frac{v}{f}\right)^2\int_0^1F^{S-HO}(x)dx,
\end{equation}
where
\begin{equation}
F^{S-HO}(x)=m_W^2\left(\frac{2}{9}f_a^{\phi^P,W_H}(x)+f_a^{\phi^{--},W_H}(x)+\frac{t_W^2}{4}f_b^{A_H,\phi^-}(x)+2f_b^{W_H,\phi^{--}}(x)
+\frac{1}{3}\int_0^{1-x}f_d(x,y) dy\right),
\end{equation}
with $f_a(x)$ and $f_b(x)$ given in Eqs. (\ref{f_a})-(\ref{f_b}), while $f_d(x,y)$, which accounts for the diagrams (c) and (d) of Figure \ref{fig_scaldiagTP}, is given by
\begin{equation}
f_d(x,y)=\frac{1}{2m_W^2\lambda(x,y) x_{W_H}}\left(2\,y\,x_{W_H} + \left( 3\,x-1 \right)\lambda(x,y) \log (\lambda(x,y) )\right),
\end{equation}
with
\begin{equation}
\lambda(x,y)=x^2 + x(x_{\phi^{--}} - x_{\phi^{-}}-1) + x_{\phi^{-}} + y(x_{W_H}-x_{\phi^-}),
\end{equation}
The respective corrections  to the $\Delta Q$ form factor are given by
\begin{equation}
\Delta Q^{S-HO}=\frac{g^2}{16\pi^2}\left(\frac{v}{f}\right)^2\int_0^1G^{S-HO}(x)dx,
\end{equation}
where
\begin{equation}
G^{S-HO}(x)=m_W^2\left(\frac{2}{9}g_a^{\phi^P,W_H}(x)+g_a^{\phi^{--},W_H}(x)+\frac{t_W^2}{4}g_b^{A_H,\phi^-}(x)+2g_b^{W_H,\phi^{--}}(x)
+\frac{1}{3}\int_0^{1-x}g_d(x,y) dy\right),
\end{equation}
with $g_a(x)$ and $g_b(x)$ given in Eqs. (\ref{g_a})-(\ref{g_b}), and
\begin{equation}
g_d(x,y)=\frac{2xy \left( 1 - x - y \right)}{m_W^2\lambda(x,y) x_{W_H}}.
\end{equation}
\subsection{Gauge boson contributions}

\begin{figure}
 \centering
\includegraphics[width=3in]{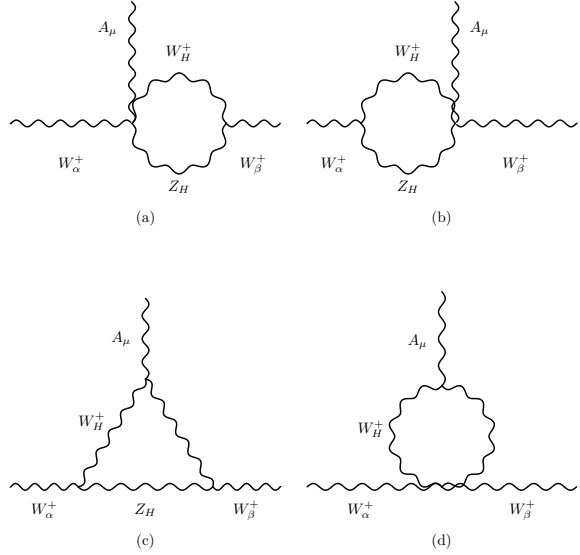}
\caption{Feynman diagrams of the gauge boson contribution
to the on-shell $WW\gamma$ vertex at the one-loop level in the LHM,
excluding SM contributions. There are four other sets  similar to
the one composed by  diagrams (a) to (c) whose contribution is multiplied by an additional suppression factor of the order of
 $(v/f)^4$. Such contributions will not be considered in our calculation. Due to the modification of the SM vertex $WWZ$, Feynman diagrams (a)-(c) also contribute when the $W_H$ and $Z_H$ gauge bosons are replaced by their SM counterparts.} \label{fig_bosdiag}
\end{figure}

We first present the type a corrections that arise from the Feynman diagrams shown in Figure \ref{fig_bosdiag}. This class of corrections is the same in both the LHM and the LHM with T-parity. There are additional contributions that include SM gauge bosons and heavy gauge bosons but apart from the suppression coming from the loop amplitude, they would receive an additional suppression factor of
of $(v/f)^4$ due to vertex corrections.  It is interesting to note that the amplitudes of all
the Feynman diagrams of Fig. \ref{fig_bosdiag} must be summed over
to cancel ultraviolet divergences. As far as gauge invariance is
concerned, bubble diagrams labeled  (a) and (b) must be added up to
render a gauge invariant amplitude, whereas diagrams labeled (c) and
(d) give a gauge invariant amplitude by themselves. The contribution
of the Feynman diagrams of Fig. \ref{fig_bosdiag} can be written
as\footnote{For  the remaining sets of Feynman diagrams described in
Fig. \ref{fig_bosdiag}, each $\hat f_{i}^{Z_H,W_H}(x)$ function must
be multiplied by an additional coefficient, analogue to those
appearing in Eqs. (\ref{F_a})-(\ref{F_c}), that arises from the
coupling constants associated with the interaction vertices involved
in each Feynman diagram. These coefficients are rather lengthy but
they are of the order of $(v/f)^4$.}

\begin{equation}
\Delta \kappa^{G}=\frac{g^2}{16 \pi^2}\sum_{i=a,c,d}\int^1_0 \hat f^{Z_H,W_H}_i(x) dx,
\end{equation}
where the $\hat f_{i}^{Z_H,W_H}(x)$ functions stand for the
contributions of the Feynman diagrams (a) to (d). The subscript $a$
denotes the sum of the contributions of the bubble diagrams (a) and
(b), which are to be added up to give gauge invariance, while the subscripts $c$
and $d$ denote the contribution of Feynman diagrams labeled (c) and
(d), respectively.  The total contribution of the Feynman diagrams
(a) and (b) is given by

\begin{eqnarray}
\hat f_{a}^{A,V}(x)&=&\frac{1}{x_A x_V}\bigg\{ \Big[2\big((x_A+x_V)x^3-(2x_A+x_V)x^2-(2x_A+1)x_Vx+(x_A+1)x_V\big)
                                           \nonumber\\
{}&{}&-\lambda \big( x_A+3x_V+1-2(2x_A+2x_V+1)x\big)\Big]\log(\lambda)-(2x-1)\lambda(x_A+x_V+1) \bigg\},
\end{eqnarray}
whereas the contribution of the Feynman diagrams labeled (c) and (d) are
\begin{eqnarray}
\hat f_c^{A,V}(x)&=&\frac{(x-1)}{2\lambda\, x_A x_V}\bigg\{ (2x_A+4x_V-1)x^5+(1-8x_A-4x_V)x^4
                                            \nonumber\\
&{}&+\Big[9x_A-6x_Vx_A -7x_V+2\lambda(2x_A+4x_V-1)+1 \Big]x^3
                                            \nonumber\\
&{}&+\Big[4x_Vx_A+x_A+7x_V-2\lambda(3x_A+4x_V-1)-1\Big]x^2
                                            \nonumber\\
&{}&+\Big[(8x_A-11x_V-4)\lambda^2 -(30x_Vx_A + 7x_A+7x_V-1)\lambda + (3-5x_A)x_V \Big]x \nonumber\\
&{}&+5\lambda^2-\lambda +3\lambda^2x_A +\lambda x_A-2\lambda^2x_V+7\lambda x_V+10\lambda x_Ax_V+3x_Ax_V -3x_V
\nonumber\\
&{}&+\lambda \Big[(10x_A+29x_V-5)x^3+(-26x_A-21x_V+3)x^2+\big(x_A(10-45x_V)+18(2\lambda-1)x_V\big)x
\nonumber\\
&{}&+ 3\big( \lambda(x_A-5x_V-1)+(5x_A+4)x_V \big) \Big]\log(\lambda) \bigg\},
\end{eqnarray}

\begin{eqnarray}
\hat f_d^{A,V}&=&\frac{3}{2}\Big(3\log(x_V)+1 \Big),
\end{eqnarray}
where $\lambda=\lambda^{A,V}(x,y)$ is given by  Eq.
(\ref{lambda}). Note that we have dropped the ultraviolet
singularities since, as explained above, they cancel when all the contributions are
summed over.

As far as $\Delta Q$ is concerned, it only receives a contribution
from the triangle diagram:

\begin{equation}
\Delta Q^{G}=\frac{g^2}{16 \pi^2}\int^1_0  \hat g^{Z_H,W_H}_c(x) dx,
\end{equation}
where
\begin{eqnarray}
\hat g_c^{A,V}(x)&=&-\frac{(x-1)^2}{3\lambda x_A x_V}\bigg[x(x-1) \Big( (2x_A+x_V-1)x^2 -2x_Ax-x_A -x_V -9x_Ax_V+1+ \lambda(2x_A+x_V-1) \Big)                                            \nonumber\\
&{}& +\lambda \Big( 5(2x_A+x_V-1)x^2-2(4x_A+x_V-1)x +x_A \Big)\log(\lambda)
\bigg],
\end{eqnarray}
with $\lambda=\lambda^{A,V}(x)$ given by Eq. (\ref{lambda}).

There are also type b contributions arising from corrections to the SM vertex $WWZ$. These are given by loops similar to those shown in Fig. \ref{fig_bosdiag}(a)-\ref{fig_bosdiag}(c) but with the heavy gauge bosons $W_H$ and $Z_H$ replaced by their SM counterparts.   The $WWZ$ is corrected up to the order of $(v/f)^4$ due to gauge boson mixing but it can get a correction of the order of $(v/f)^2$ due to the modification of
$G_F$ by $W_H$ exchange \cite{Han:2003wu}. The respective contributions to the $W$ form factors are given by

\begin{equation}
\Delta \kappa^{G}_{SM}=\frac{g^2}{16 \pi^2}h^{Z,W}\left(\frac{v}{f}\right)^2\sum_{i=a,c}\int^1_0 \hat f^{Z,W}_i(x) dx,
\end{equation}
\begin{equation}
\Delta Q^{G}_{SM}=\frac{g^2}{16 \pi^2}h^{Z,W}\left(\frac{v}{f}\right)^2\int^1_0 \hat g^{Z,W}_a(x) dx,
\end{equation}
with $h^{Z,W}=\frac{c_W^2}{4\cos 2\theta_W} \left(-4c^2s^2 + 5(c'^2 - s'^2)^2 \right)$
\section{Numerical results and analysis}
\label{Analysis}

\subsection{Constraints on the model parameters}
Several parameters of the LHM are involved in our calculation, such
as the scale of the global symmetry breaking $f$, the mixing angle
of the gauge sector $\tan \theta=s/c=g_1/g_2$, the VEV of the
$SU(2)$ triplet $v'$, etc. Constraints on these parameters have been
obtained from electroweak precision measurements
\cite{Csaki:2002qg,Hewett:2002px,Csaki:2003si,Gregoire:2003kr,Chen:2003fm,Casalbuoni:2003ft,Marandella:2005wd,Han:2005dz,
Kilian:2003xt, Hubisz:2005tx}, including the $Z$ pole data,
low-energy neutrino-nucleon scattering, and the $W$ mass
measurement.  In the LHM, weak isospin violating contributions to
electroweak precision observables are present as there is no
custodial $SU(2)$ global symmetry \cite{Csaki:2002qg}. The largest
corrections are due to the exchange of the heavy gauge bosons
\cite{Csaki:2002qg}.  A global fit was performed to experimental
data \cite{Csaki:2002qg} and it was found that the symmetry breaking
scale $f$ is severely constrained throughout most of the parameter
space: $f
> 4$ TeV at 95\% C.L. \cite{Csaki:2002qg} for generic values of the mixing parameters
$c$ and $c'$. In the best scenario, a suitable election of the
hypercharge assignments of the light fermions under the two $U(1)$
factors allows for $f\sim 1$ TeV for a small region of the parameter
space \cite{Csaki:2003si}. Although the LHM seems tightly
constrained, it was suggested in Ref. \cite{Han:2003wu} that
dangerous corrections to electroweak precision observables may be
controlled by tuning the parameters of the model, which would allow
for less stringent constraints on the scale $f$.  This tuning  seems
however unnatural and contrary to the spirit of the model. As
commented before, the addition of the discrete symmetry T-parity can
alleviate this problem \cite{Low:2004xc} by forbidding any
interaction induced by the triplet VEV $v'$ and the tree-level
contributions to electroweak observables arising from the heavy
gauge bosons. Corrections to electroweak observables are generated
up to the one-loop level and constraints on the model parameters are
significantly weaker than in the original LHM, allowing $f$ to be as
low as 500 GeV \cite{Hubisz:2005tx}.

Below we will examine the size of the different contributions to the
$\Delta \kappa$ and $\Delta Q$ form factors as a function of the scale $f$.   To analyze the behavior of the form factors, we will consider the region 500 GeV $\le f \le$ 6000 GeV. The allowed region can be taken as $f> 4$ GeV in the LHM, whereas the allowed region in the LHM with T-parity is $f>500$ GeV.  To be consistent with electroweak precision constraints we will use $c'=s'$, which in fact  is true in the LHM with T-parity.  As far as the
mixing parameter $c$ is concerned,  we consider the range $0.1\le c\le 0.9$ which is required for the $SU(2)_H$ coupling not to
be strongly interacting. Although the lower bound on $f$ depends on the actual value of $c$, our analysis and conclusions will remain unchanged, as $f>4\sim 5$ TeV in the region considered for $c$. We also would like to stress that the value $c=s$ corresponds to the LHM with T-parity, but it also applies  to the LHM without T-parity as in such a case the calculated contributions to the $W$ form factors coincide. For the masses of the heavy particles, we  will use the lower bound on the scalar triplet mass given by Eq. (\ref{Phimasslimit}), whereas for the heavy gauge boson masses we will use the exact expressions given in Eq. (\ref{WHmasslimit}) and (\ref{AHmasslimit}). The new contributions to the $W$ form factors will be given in units of the constant $a=g^2/(96\pi^2)$, which has been  customarily used in other calculations of the $W$ form factors reported in the literature \cite{Bilchak:1986bt,Couture:1987xq,Lahanas:1994dv,Aliev:1985an,Couture:1987eu, TavaresVelasco:2001vb,GarciaLuna:2003tj}.

\subsection{Fermion contribution}

\begin{figure}
 \centering
\includegraphics[width=3in]{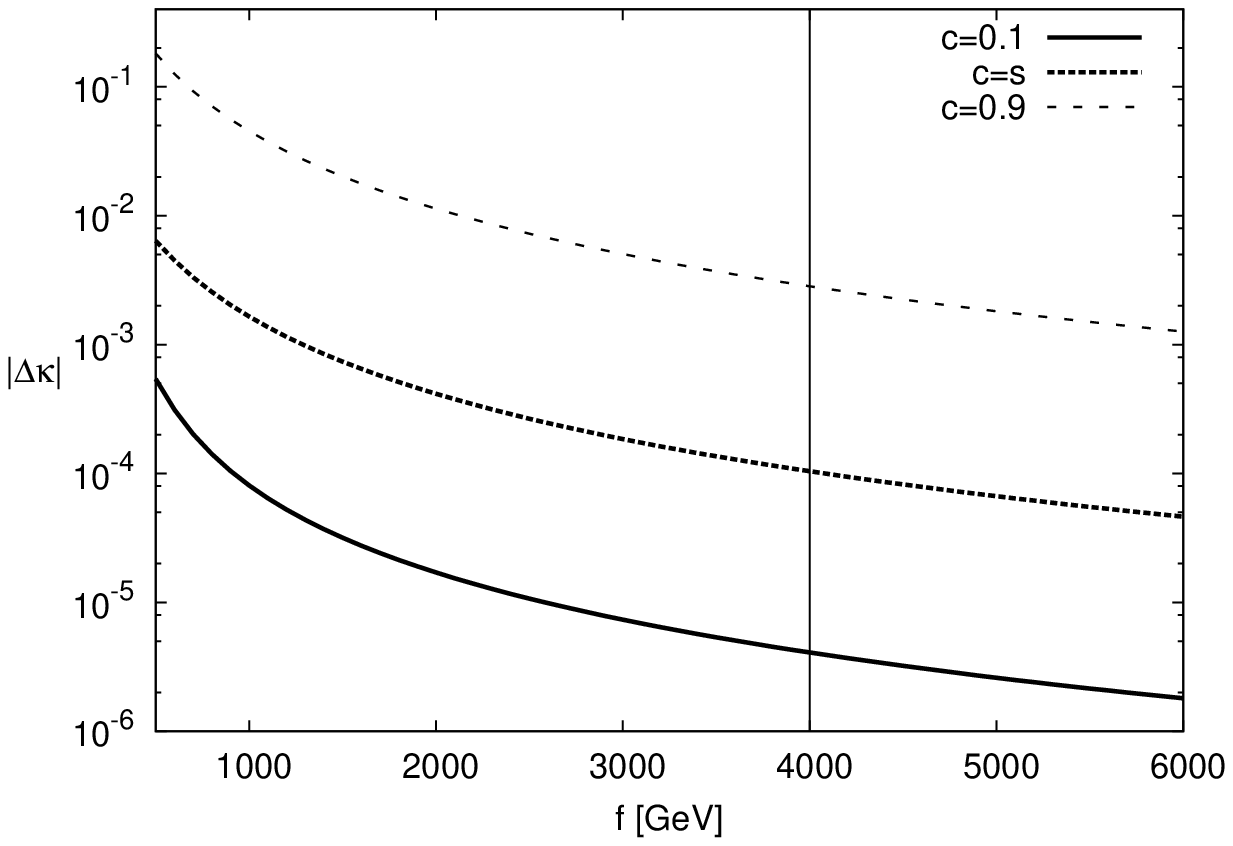}\includegraphics[width=3in]{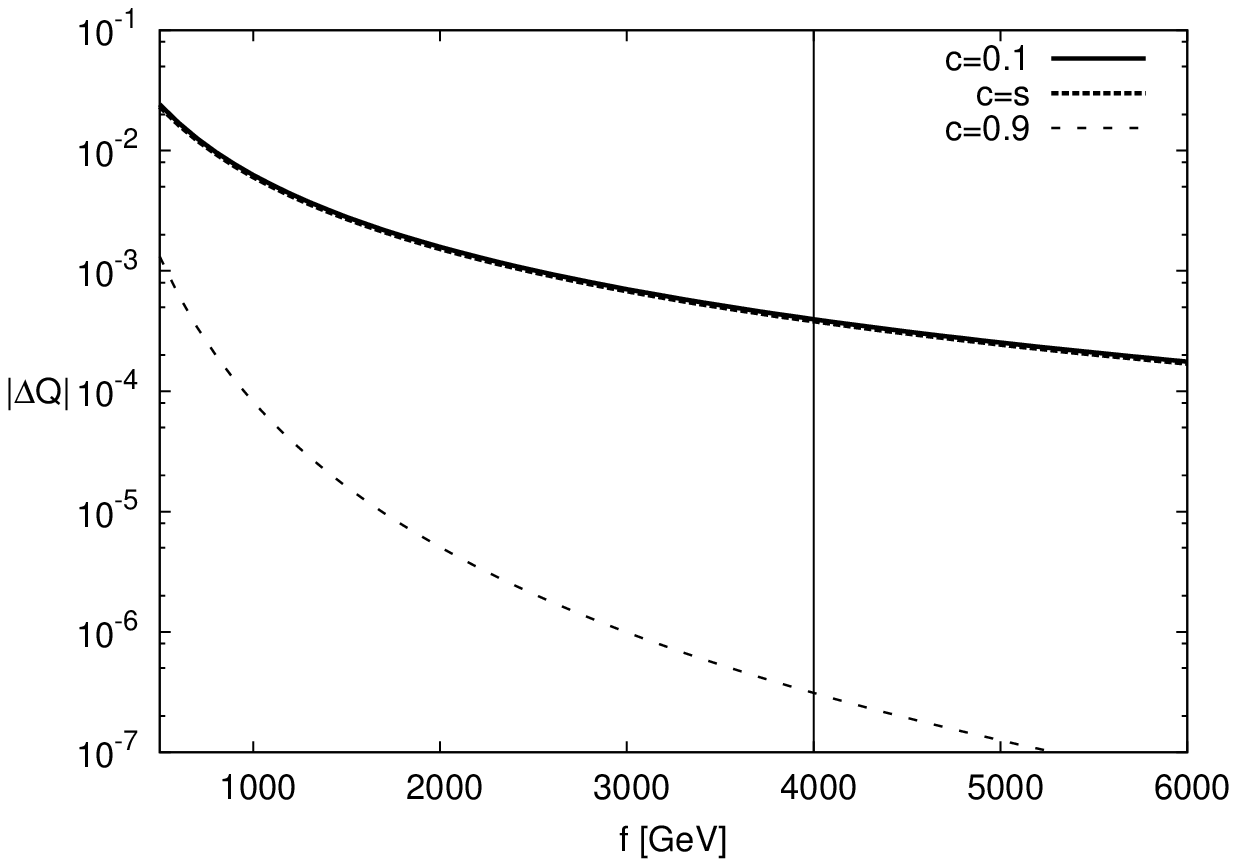}
\caption{Fermion contribution to the $W$ form factors in the LHM  as a function of the scale  $f$ for two values of the mixing parameter $c$  and $x_L=1/2$, in which case  $m_{T,t'_+}=\sqrt{2} f$. The curves for $c=s$ correspond to the LHM with T-parity. To the right of the vertical line  there is the allowed region of the LHM without T-parity, but the complete region is still allowed in the LHM without T-parity.}
\label{ferKaCon}
\end{figure}

The corrections to the $W$ form factors from the fermion sector depends on the parameters  $c$, $x_L$, and $f$.  We will use the value $x_L=1/2$, which corresponds to a mass of the heavy top partner given by $m_T=\sqrt{2}f$. We will  analyze the dependence of the form factors  for three values of the mixing parameter $c$, namely, $c=0.1$, $c=s$, and $c=0.9$. The results are shown in Fig. \ref{ferKaCon}. The full $f$ region shown in the plots is allowed in the LHM with T-parity but only the region to the right of the vertical line is the one allowed in the LHM without T-parity.  We can observe that the size of $\Delta \kappa$ is larger  for $c=0.9$ and smaller for $c=0.1$, while $\Delta Q$  shows the opposite behavior, though it  is highly suppressed for $c=0.9$. For $c=s$ and $c=0.1$ the curves for $\Delta Q$ are almost indistinguishable.

\subsection{Scalar boson contribution}
In the LHM, the mixing parameters of the scalar sector
can be approximated as $s_0\simeq 2\sqrt{2}v'/v$ and $s_+\simeq
2 v'/v$ \cite{Han:2003wu}. In this limit the coefficients $h^{\phi^0,W_H}$ and $h^{\phi^0, W}$ vanish. Also, those Feynman diagrams whose amplitude is proportional to $v'^2$ are suppressed by an additional factor of $(v'/m_W)^2$. Since, constraints from electroweak precision observables indicate that $v'\ll v$, we will neglect also those contributions, which exactly cancel in the LHM with T-parity indeed. Our calculation consider thus the following corrections to the $W$ form factors. First, there is the type a contribution from the Feynman diagram of Fig. \ref{fig_scaldiag}(a) with
the light Higgs boson $H$ and the heavy gauge boson $W_H$ running in
the loop, which is proportional to $(c^2-s^2)^2$ and it could give a non-negligibly contribution for values of the mixing parameter $c$ such that $(c^2-s^2)^2\sim O(1)$. This contribution however vanishes in the LHM with T-parity. Also, we will include the type a contribution from the loops with only heavy scalars, and the type b and type c contributions described above. The latter can be non-negligibly  for $f\lesssim 1$ TeV.

\begin{figure}
 \centering
\includegraphics[width=3in]{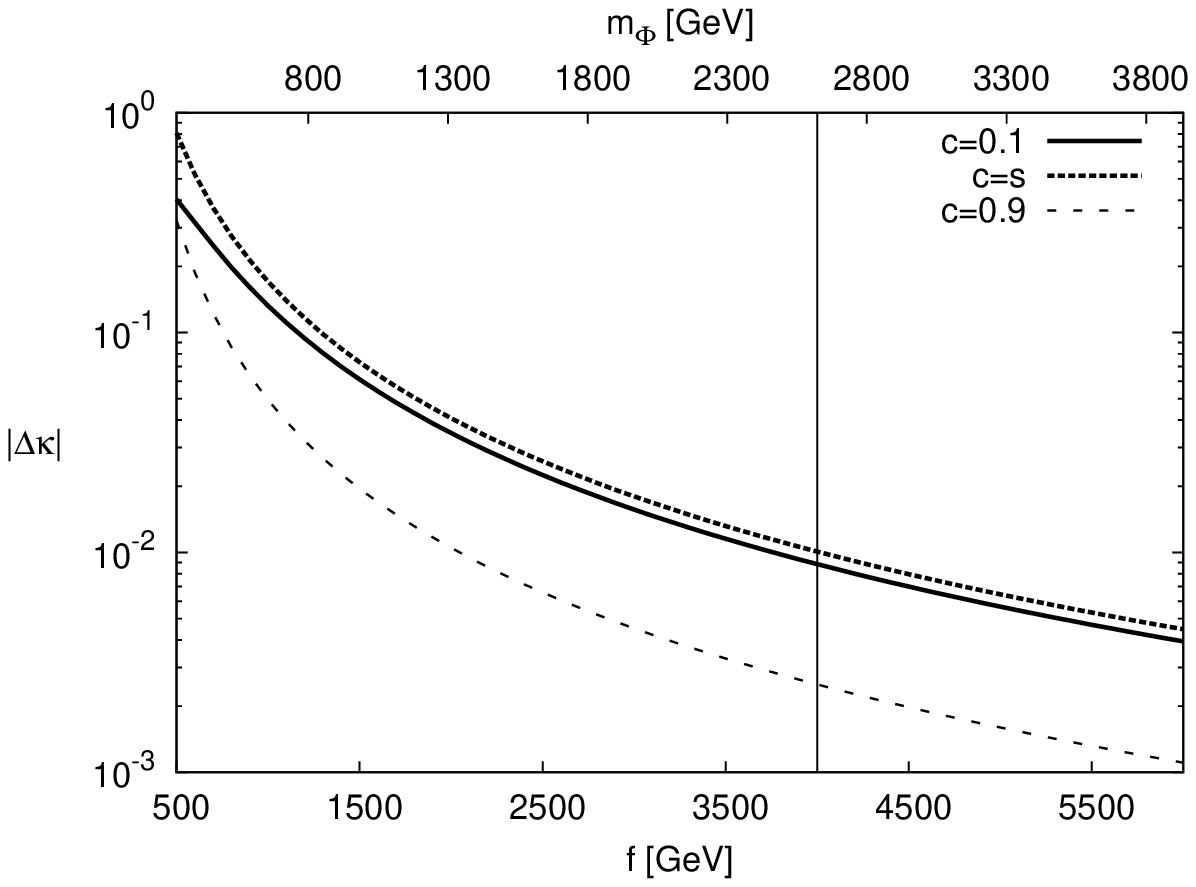}\includegraphics[width=3in]{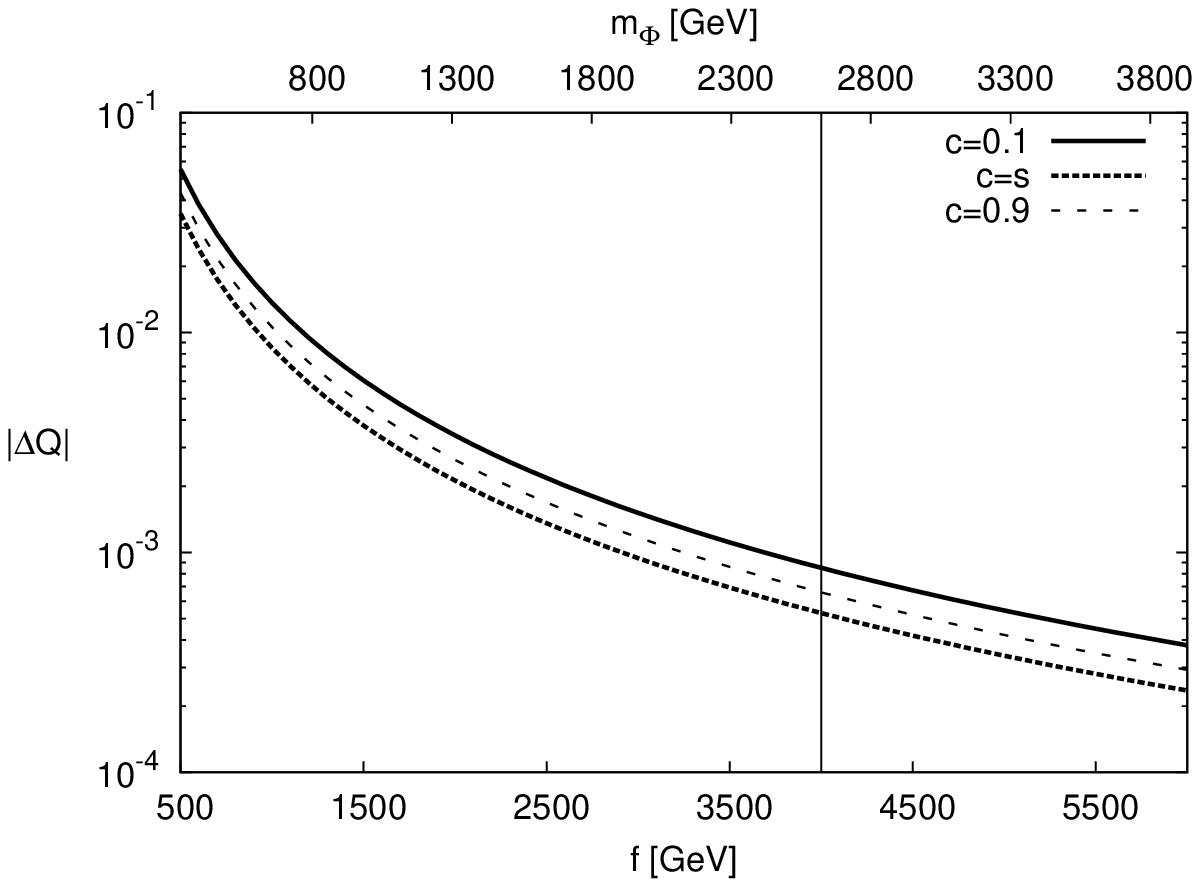}
\caption{Scalar contribution to the $W$ boson form factors in the LHM, in units of
$a=g^2/(96\pi^2)$ as a function of the scale $f$ and for three values of the mixing parameter $c$. The curve for $c=s$ corresponds to the LHM with T-parity. The mass of the scalar triplet is set to the lower bound $m_\Phi=\sqrt{2}m_H f/v$ with $m_H=114$ GeV, whereas the masses of the heavy gauge bosons are given in Eqs. (\ref{WHmasslimit}) and (\ref{AHmasslimit}).  The allowed region for the $f$ parameter due to electroweak constraints lies to the right from the vertical line in the LHM, but the whole region is still allowed in the LHM with T-parity.} \label{ScaKaT}
\end{figure}

  In Fig. \ref{ScaKaT} we show the behavior of the scalar contribution to $\Delta \kappa$  and $\Delta Q$ as a function of the scale $f$ for three representative values of the mixing parameter $c$. The dominant contribution to $\Delta \kappa$ arises from the loop with the gauge boson $W$ and the Higgs boson $H$ (type b contribution). Since the contribution from the loop with the heavy gauge boson $W_H$ and the Higgs boson $H$ is proportional to $(c^2-s^2)^2$, it reaches its large size when $(c^2-s^2)^2\sim O(1)$. In this case this contribution is of the same order of magnitude than the type b contribution. However, for $c=0.1$ these contributions have opposite sings and their sum can suffer from cancelations. The remaining contributions to $\Delta \kappa$ (type a and type c) are subdominant, with the type c contribution being important only for $f\lesssim 1$ TeV. As far as $\Delta Q$ is concerned, it gets the dominant correction from the loops with heavy scalars (type a contribution) and the type b contribution. The type a contribution from the Feynman diagram of Fig. \ref{fig_scaldiag}(a) is subdominant and so are the type c contributions. As a result, $\Delta Q$ is almost independent of the value of the mixing parameter $c$. In summary, $\Delta \kappa$ is larger for $c=s$ and lower for $c=0.9$, whereas  $\Delta Q$ is almost indistinguishable in the LHM and the LHM with T-parity. Both form factors decrease quickly for increasing $f$.

\subsection{Gauge boson contribution}
\begin{figure}
 \centering
\includegraphics[width=3in]{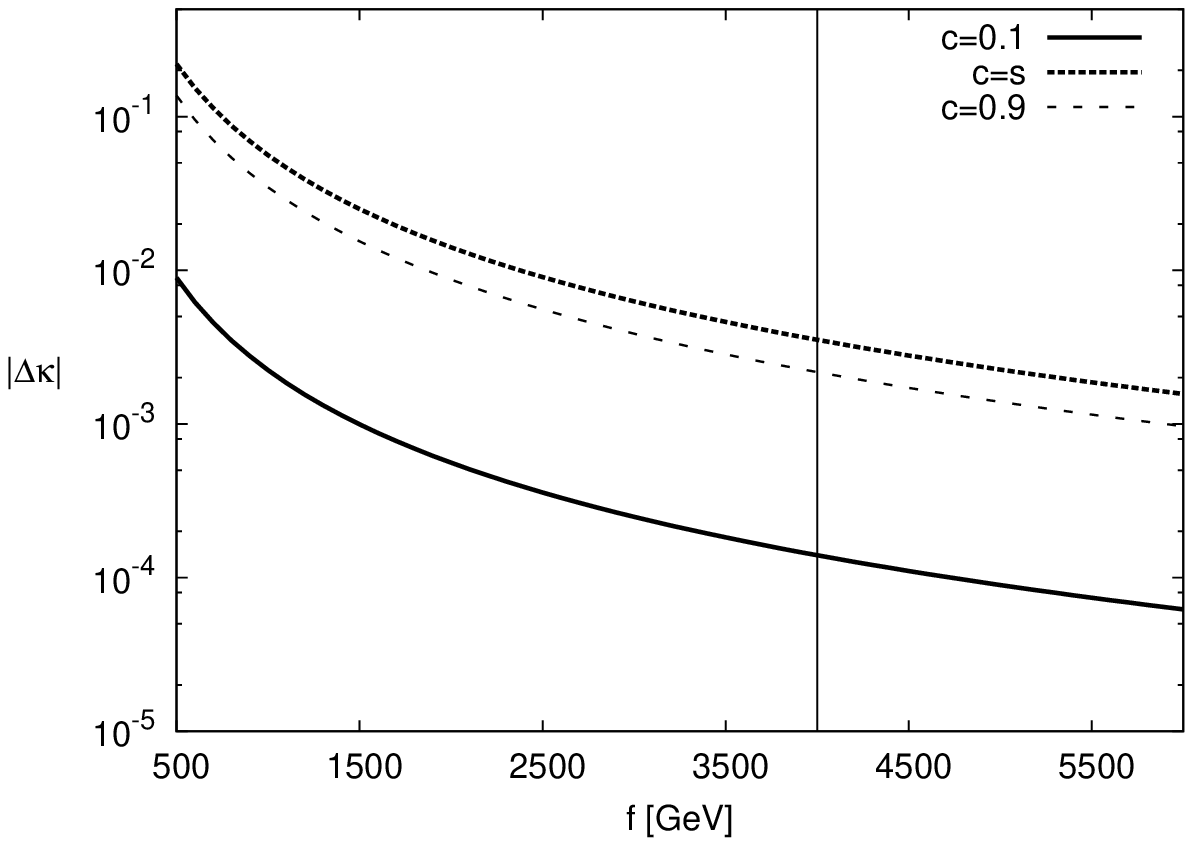}\includegraphics[width=3in]{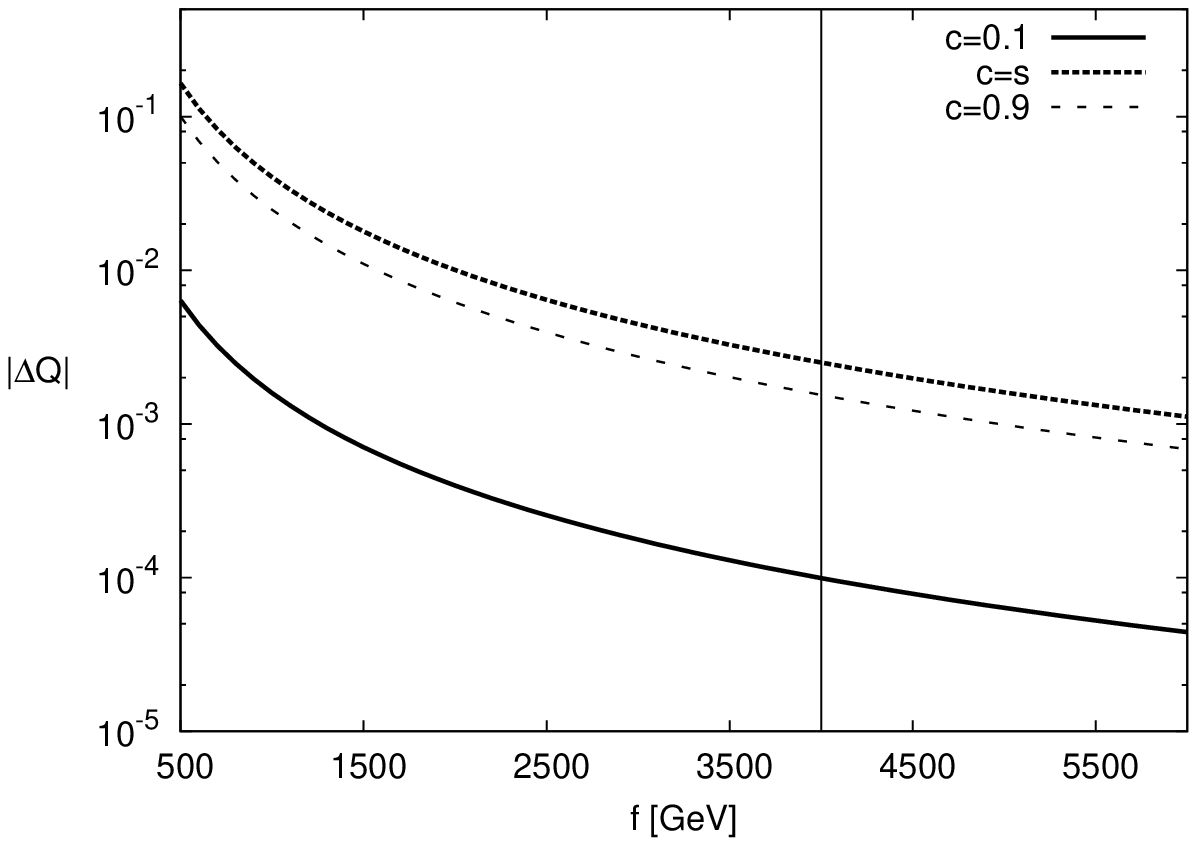}
\caption{Gauge boson contribution to the $W$ gauge boson form
factors  in the LHM, in units of $a=g^2/(96\pi^2)$, as a function of the scale $f$ and for three values of the mixing parameter $c$. The value $c=s$ corresponds to the LHM with T-parity. The  masses of the heavy gauge bosons are given in Eq. (\ref{WHmasslimit}).  The allowed region for the $f$ parameter due to electroweak constraints lies to the right from the vertical line in the LHM, but the whole region is still allowed in the LHM with T-parity.} \label{BosConFF}
\end{figure}

The gauge boson corrections to  $\Delta \kappa$ and $\Delta Q$ are shown in Fig.
\ref{BosConFF} as a function of the scale $f$. The  masses of the heavy gauge bosons are given in Eq. (\ref{WHmasslimit}).
The dominant gauge boson contribution arises from the type b contribution whereas the Feynman
diagrams of Fig. \ref{fig_bosdiag}  (type a contribution) are subdominant. For small values of $c$, the gauge boson contributions are very suppressed as the heavy gauge boson masses ${m_W}_H$ and ${m_Z}_H$ are very large for a fixed $f$. In general, both  $\Delta \kappa$ and $\Delta Q$ receive very small corrections for very large $f$. The gauge boson contributions to the $W$ form factors have a magnitude similar
to those arising from the fermion sector and  from the Feynman diagrams with only heavy scalar particles [Fig. \ref{fig_scaldiag}(c)], but all these contributions can add destructively or constructively depending on the value of $c$.

\subsection{Total contribution}
Finally, we  present the total contribution to the $W$ form factors from the LHM  as a function of the scale $f$ for the same values of the parameters of the model used in the above analysis. The dominant contributions to the $W$ form factors are those of type a and type b, whereas the type c contributions are only important for $f< 1$ TeV. However, the partial contributions can add constructively or destructively depending on the value of the mixing parameter $c$. For values of $f\lesssim 1$ TeV, the effects of the higher order corrections can be important as the factor $v/f$ is not very suppressed. When $c=0.1$ and $c=s$, there is no cancelation between the partial contributions to  $\Delta \kappa$, which can reach its larger size. As far as $\Delta Q$ is concerned, the partial contributions add destructively for $c=0.1$ but they add constructively for $c=s$ and $c=0.9$.

 In summary, in the allowed region of the LHM without T-parity ($f>4$ TeV), the new contributions to the $W$ form factors are very suppressed. However, for $f$ around 500 GeV, a region still allowed in the LHM with T-parity, $\Delta \kappa$ and $\Delta Q$  can be of the order of $10^0-10^{-1}$, in units of the constant $a$. To have an idea of the size of these corrections, the $W$ form factors are corrected in the SM by terms of the order of: $\Delta \kappa\sim 10a$ and $\Delta Q\sim a$ \cite{WWgSM}. The contribution from other weakly coupled SM extensions are about one percent of the SM corrections for values of the masses of the new particles consistent with the current experimental bounds \cite{Bilchak:1986bt,Couture:1987xq,Lahanas:1994dv,Aliev:1985an,Couture:1987eu,TavaresVelasco:2001vb,GarciaLuna:2003tj}. After inserting the value of the constant $a$, we conclude that the largest corrections to the $W$ form factors are obtained in the LHM with T-parity: $|\Delta \kappa|\sim 4.5 \times 10^{-4}$ and $|\Delta Q|\sim 7.7 \times 10^{-5}$ for $f=500$ GeV. We will see below that these values may get some enhancement from the strongly-coupled ultraviolet (UV) completion of the LHM.

\begin{figure}
 \centering
\includegraphics[width=3in]{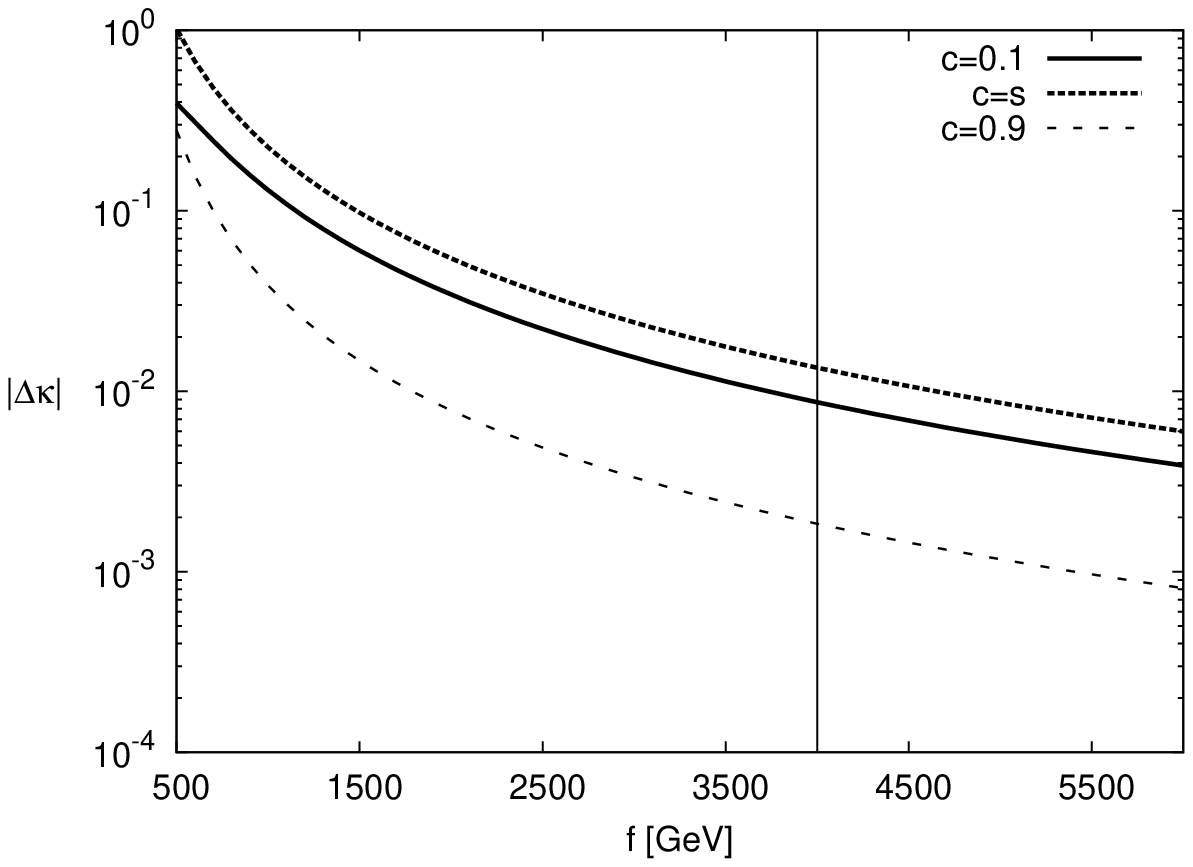}\includegraphics[width=3in]{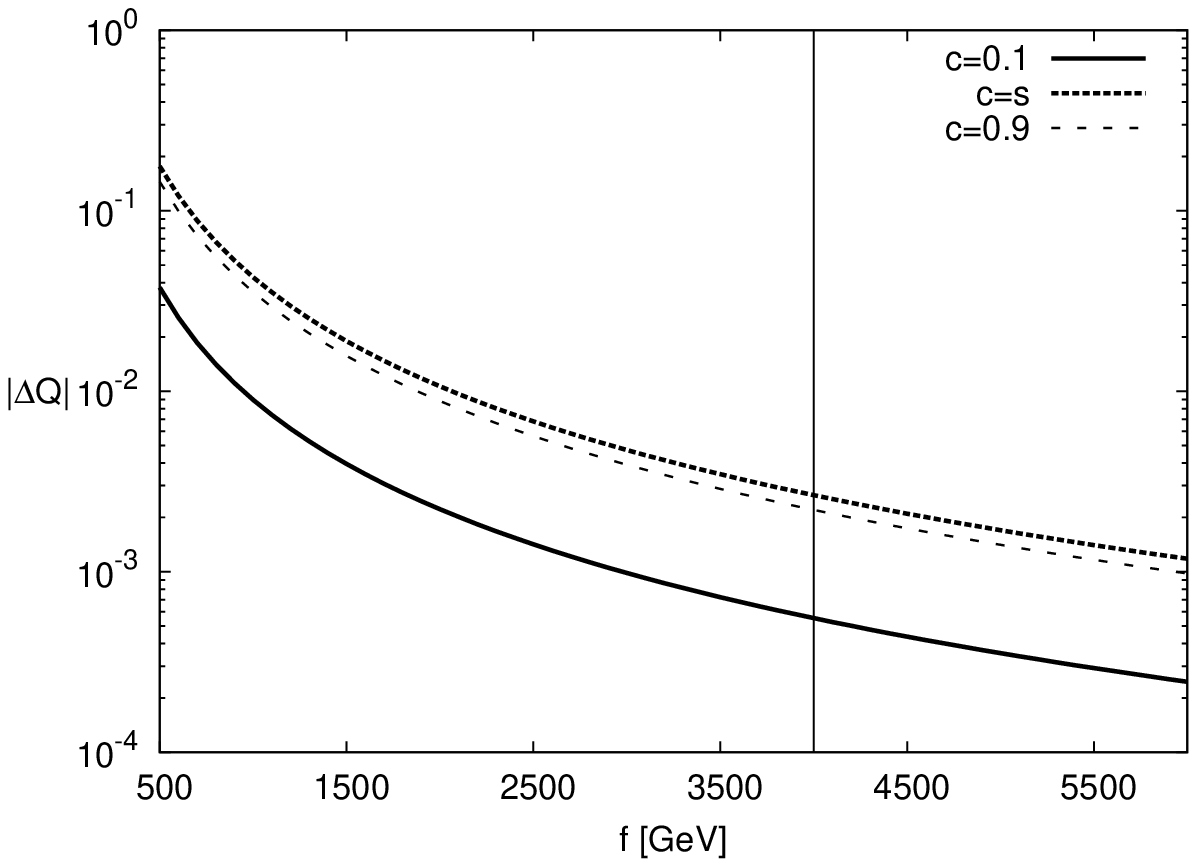}
\caption{One-loop corrections  to
$\Delta \kappa$ and $\Delta Q$ in the LHM for three
values of the mixing parameter $c$, in units of $a=g^2/(96\pi^2)$, as a
function of the scale $f$. The curve for $c=s$  corresponds to the LHM with T-parity.  For the masses of the heavy scalars we used the lowest bound $m_\Phi=\sqrt{2} m_H f/v$ with $m_H=114$ GeV,  the masses of the heavy gauge bosons are given in Eqs. (\ref{WHmasslimit}) and (\ref{AHmasslimit}), and the mass of the heavy top partner was taken as $m_T=\sqrt{2}f$, which corresponds to $x_L=1/2$.  The allowed region for the $f$ parameter due to electroweak constraints lies to the right from the vertical line in the LHM, but the whole region is still allowed in the LHM with T-parity.} \label{ConTotKa}
\end{figure}

\subsection{Estimate of the contribution of the ultraviolet completion of the LHM to the $W$ form factors}
The LHM is an effective theory valid up to the cutoff scale $\Lambda=4 \pi f$. Above this scale the physics becomes strongly coupled,
and the LHM model must be replaced by its UV completion, which would be a QCD-like gauge theory with a confinement scale around 10 TeV \cite{ArkaniHamed:2002qy,Katz:2003sn,Thaler:2005en}. This gives rise to the possibility that EWSB is driven by  strong dynamics such as occurs in technicolor theories. Although we will not give a detailed account of the features of such an UV completion, it would be interesting to consider the possibility that the $W$ form factors can receive important corrections from strongly interacting couplings. An example of such UV completions is that discussed in \cite{Katz:2003sn}, where
a slightly modified version of the LHM was embedded into a renormalizable supersymmetric theory valid up to scales
of the order of the Planck scale. This UV completion considers a new ultracolor gauge interaction, with the ultracolor gauge group $SO(7)$, and a set of ultrafermions transforming in a real representation of $SO(7)$.

 An estimate for the size of the $W$ form factors from strongly interacting theories such as technicolor was already obtained via the effective lagrangian approach and na\"ive dimensional analysis in Ref. \cite{Wudka}. These results are rather general and can be used to get an estimate of the size of the contributions of the LHM UV completion, which may arise from a technofermion loop for instance. The parametrization used  for the CP-conserving $WW\gamma$ coupling is the one given in Eq. (\ref{WWgHagi}). In order to estimate the size of the electromagnetic $W$ form factors due to strongly interacting couplings, the effective Lagrangian  can be written as the one of a gauged chiral model and the appropriate expansion of the Lagrangian is given in  four-momentum powers. Though strongly coupled, the effective Lagrangian must represent a consistent theory at low energies, which means that the tree-level contributions are  not to be swamped by radiative corrections. In the chiral effective Lagrangian $\kappa_\gamma$ is associated with a dimension-four operator while $\lambda_\gamma$ is associated with a dimension-six operator. Na\"ive dimensional analysis allows one to obtain the following estimates\cite{Wudka}:

\begin{eqnarray}
|\Delta\kappa_\gamma|&\sim& \frac{1}{(4\pi)^2}\sim 6\times 10^{-3},\\
|\lambda_\gamma|&\sim& \frac{g^2}{(4\pi)^4}\sim 10^{-5},
\end{eqnarray}

  This translates into the estimate $|\Delta\kappa|\sim 6.2\times 10^{-3}$ and $|\Delta Q|\sim 2\times 10^{-5}$, which agrees with the estimate of the size of the $W$ form factors that was obtained within the framework of technicolor theories in Ref. \cite{Appelquist:1993ka}, where the authors considered the contribution to the $WW\gamma$ vertex from a heavy fermion $SU(N)$ doublet. These strongly interacting contributions would enhance  the size of the $W$ form factors in the LHM unless the confinement scale is too large.

\section{Final remarks}
\label{Remarks}

We have calculated the one-loop corrections to the static electromagnetic properties of the $W$ gauge boson in both the LHM
and the LHM with T-parity. We also considered an estimate of the contribution of the strongly coupled UV completion of the theory. As far as the corrections from the weakly coupled interactions are concerned, the new contributions depend on the mixing parameter $c$ and the scale $f$. The partial contributions can add constructively or destructively depending on the value of $c$. We considered the contributions up to the order of $(v/f)^4$, which are subdominant but can add constructively to enhance somewhat the $W$ form factors for values of $f\lesssim 1$ TeV.
  As a matter of fact, the corrections to the $W$ form factors are expected to be very suppressed in the LHM without T-parity since the scale $f$ is strongly constrained in this version of the model. The largest contribution to both form factors would arise in the LHM with T-parity ($c=s$)  for $f$ around 500 GeV, which is a region still allowed in this model. The size of such corrections is about the same order of magnitude than the corrections of other (weakly coupled) extensions of the SM, such as supersymmetry
theories \cite{Bilchak:1986bt,Couture:1987xq,Lahanas:1994dv,Aliev:1985an},
the two-Higgs doublet model \cite{Couture:1987eu}, 331 models
\cite{TavaresVelasco:2001vb,GarciaLuna:2003tj}. However, these corrections seem to be below the level of experimental sensitivity in future measurements at the LHC and  the planned ILC. There is however the possibility that the strongly interacting UV completion of the LHM may enhance the size of the $W$ form factors by about one order of magnitude.

\acknowledgments{Support
from SNI (Mexico) and VIEP-BUAP is acknowledged. We would like to thank I. Fuentecilla for participating in the early stage of this work.}

\appendix
\section{Feynman rules}
\label{FeynRules} In this appendix we present the Feynman rules
necessary for our calculation. In Table \ref{Tab_Feynrules} we show
the Feynman rules for the LHM \cite{Han:2003wu}.
\begin{table}
\begin{tabular*}{0.4\textwidth}{@{\extracolsep{\fill}}ll}
\hline\\
Vertex&Feynman rule\\
\hline
$ \phi^+(p_1)\phi^-(p_2)A_\mu$&$-ieR_\mu$\\

$\phi^{++}(p_1)\phi^{--}(p_2)A_\mu $&$-2ieR_\mu$\\

$W^+_\mu H(p_1)\phi^-(p_2)$&$-\frac{i g}{2}(\sqrt{2}s_0-s_+)R_\mu$\\

$W^+_\mu \phi^0(p_1)\phi^-(p_2)$&$-\frac{i g}{\sqrt{2}}R_\mu$\\

$W^+_\mu \phi^P(p_1)\phi^-(p_2)$&$\frac{g}{\sqrt{2}}R_\mu$\\

$W^+_\mu\phi^+(p_1)\phi^{--}(p_2)$&$-igR_\mu$\\

$W^+_\mu W^-_\nu \phi^{0}$&$-\frac{i g^2}{2}(s_0v-2\sqrt{2}v') g_{\mu\nu}$\\

$W^+_\mu W^+_\nu\phi^{--}$&$2ig^2 v' g_{\mu\nu}$\\

$W^+_\mu W^-_\nu H$&$\frac{ig^2v}{\sqrt{2}}\delta_{WWH}g_{\mu\nu}$\\

$W^+_\mu{W_H^-}_\nu H$&$-\frac{i g^2}{2}\frac{(c^2-s^2)v}{2sc}g_{\mu\nu}$\\

$W^+_\mu{W_H^-}_\nu \phi^{0}$&$\frac{i g^2}{2}\frac{(c^2-s^2)(s_0v-2\sqrt{2}v')}{2sc}g_{\mu\nu}$\\

$W^+_\mu {W_H^+}_\nu \phi^{--}$&$2ig^2\frac{(c^2-s^2)v'}{2sc}g_{\mu\nu}$\\

$W^+_\mu{Z}_\nu\phi^{-}$&$-\frac{i g^2v'}{c_W}g_{\mu\nu}$\\

$W^+_\mu {Z_H}_\nu\phi^{-}$&$ig^2\frac{(c^2-s^2)v'}{2sc}g_{\mu\nu}$\\

$W^+_\mu {A_H}_\nu\phi^{-}$&$\frac{igg'}{2}\frac{(c'^2-s'^2)(v{s_+}-4v')}{2s'c'}g_{\mu\nu}$\\

$W^+_\mu(k_1)W^-_\nu(k_2)A_\rho(k_3)$&$ieS_{\mu\nu\rho}$\\

${W_H}^+_\mu(k_1){W_H^-}_\nu(k_2)A_\rho(k_3)$&$ieS_{\mu\nu\rho}$\\

${W}^+_\mu(k_1){W_H^-}_\nu(k_2){Z_H}_\rho(k_3)$&$igS_{\mu\nu\rho}$\\

$A_\mu{Z_H}_\nu{W}^+_\rho{W_H^-}_\sigma$&$igeT_{\mu\nu\rho\sigma}$\\

$W^+_\mu{W_H^+}_\nu {W}^-_\rho{W_H^-}_\sigma$&$-ig^2T_{\mu\nu\rho\sigma}$\\

$W^+_\mu\bar{t}b$&$\frac{ig}{\sqrt{2}}\delta_{Wtb}\gamma^\mu P_LV_{tb}^{\rm SM}$\\

$W^+_\mu\bar{T}b$&$\frac{g}{\sqrt{2}}\frac{v}{f} x_L\gamma^\mu P_LV_{tb}^{\rm SM}$\\
\hline
\end{tabular*}
\caption{Feynman rules in the LHM \cite{Han:2003wu} that induce the
$W$ form factors. All the momenta are outgoing and we have dropped
the $L$ subscript in the light (SM) gauge bosons. $g'=g/c_W$,
$R_\mu=(p_1-p_2)_\mu$, $S_{\mu\nu\rho}=g_{\mu\nu}(k_1- k_2)_\rho +
g_{\nu\rho}(k_2 - k_3)_\mu + g_{\mu\rho}(k_3 - k_1)_\nu$, and
$T_{\mu\nu\rho\sigma}=2g_{\mu\nu}g_{\sigma\rho}-g_{\mu\rho}g_{\nu\sigma}-g_{\mu\sigma}g_{\nu\rho}$. Also $P_L=\frac{1}{2}(1-\gamma^5)$, $\delta_{WWH}=1-\frac{v^2}{f^2} (\frac{1}{3}-\frac{1}{2}(c^2-s^2)^2)$ and $\delta_{Wtb}=1-\frac{v^2}{2f^2} (x_L^2-c^2(c^2-s^2))$.
\label{Tab_Feynrules}}
\end{table}

The Feynman rules to get the corrections to the $W$ form factors in
the LHM with T-parity  can be obtained
from Table \ref{Tab_Feynrules} after the replacements of Eqs. (\ref{RepTPar1})-(\ref{RepTPar3}) are
done \cite{Hubisz:2004ft,Chen:2006cs}.

For the type c corrections to the $W$ form
factors we also need the Feynman rules given in Table
\ref{Tab_FeynrulesTP}.

\begin{table}
\begin{tabular*}{0.4\textwidth}{@{\extracolsep{\fill}}ll}
\hline\\
Vertex&Feynman rule\\
\hline
${W_H}^+_\mu W^-_\nu \phi^P$&$\frac{g^2}{3\sqrt{2}}\frac{v^2}{f}g_{\mu\nu}$\\

${W_H}^+_\mu W^+ _\nu \phi^{--}$&$-i\frac{g^2}{2}\frac{v^2}{f}g_{\mu\nu}$\\

${W_H}^+_\mu A_\nu\phi^-$&$-i\frac{eg}{6}\frac{v^2}{f}g_{\mu\nu}$\\

${W_H}^+_\mu {A_H}_\nu\phi^-$&$-i\frac{gg'}{4}\frac{v^2}{f}g_{\mu\nu}$\\

$W^+_\mu\bar{t}b$&$\frac{ig}{\sqrt{2}}\left(1-\frac{v^2}{2f^2} c_\lambda^4\right)\gamma_\mu P_LV_{tb}^{\rm SM}$\\

$W^+_\mu\bar{t'_+}b$&$-\frac{ig}{\sqrt{2}}\frac{v}{f} c_\lambda^2\gamma_\mu P_LV_{tb}^{\rm SM}$\\
\hline
\end{tabular*}
\caption{Feynman rules necessary  to calculate
the contributions to the $W$ form factors up to the
order of $(v/f)^4$ \cite{Hubisz:2004ft,Chen:2006cs}. \label{Tab_FeynrulesTP}}
\end{table}

\end{document}